\documentclass[%
 reprint,
 amsmath,amssymb,
 aps,
]{revtex4-2}

\usepackage{dcolumn}
\usepackage{bm}

\usepackage{pgfplots}
\usepgfplotslibrary{groupplots,dateplot}
\usepackage{pgf} 
\pgfplotsset{compat=newest}
\usetikzlibrary{patterns,shapes.arrows}
\usepackage{standalone}
\usepackage{graphicx}
\usepackage[justification=Justified]{caption}

\usepackage{subcaption}

\usepackage{hyperref}

\begin{document}
\preprint{APS/123-QED}

\title{Threshold for loss of Landau damping in double-harmonic rf systems}

\author{Leandro Intelisano}
\email{leandro.intelisano@cern.ch}

\author{Heiko Damerau}

\author{Ivan Karpov}%
 
\affiliation{CERN, CH 1211 Geneva 23, Switzerland}%

\date{\today}

\begin{abstract}

Landau damping is a natural stabilization mechanism that mitigates coherent beam instabilities.
In the longitudinal plane, loss of Landau damping~(LLD) occurs when a coherent mode of oscillation emerges from the incoherent band of the bunch synchrotron frequencies.
This work extends the recent LLD studies to the relevant case of double-harmonic rf systems. Specifically, it is shown that in the bunch shortening mode (both rf systems in phase at the bunch position for a non-accelerating bucket), inductive impedance above transition energy results in a vanishing LLD threshold for a binominal particle distribution, similar to the single-harmonic rf case.
In this configuration, refined analytical estimates of the synchrotron frequency distribution enabled the derivation of an analytical equation for the LLD threshold by introducing an upper cutoff frequency to the impedance.
The LLD threshold is studied through the concept of van Kampen modes and takes into account the effect of the voltage ratio, as well as the relative phase between the two rf systems for an inductive impedance above transition energy (or capacitive below).
The validity of the theoretical studies is supported by extensive beam measurements conducted under different bucket-filling conditions in two synchrotrons, the PS and the SPS at CERN.
Beyond the analytical estimates, the observations are moreover compared with the semi-analytical code MELODY and macroparticle tracking simulations in BLonD.

\end{abstract}

\maketitle

\section{INTRODUCTION}
Landau damping~\cite{Landua} is a natural mechanism that provides beam stability to a wide range of working high-intensity accelerators.
Hence, acquiring a comprehensive understanding and making accurate predictions of the threshold for the loss of Landau damping (LLD) is essential for both single and multi-bunch instabilities in current as well as future projects.
With reference to the longitudinal plane, Landau damping is established by means of the spread of synchrotron frequencies of individual particles in a bunch, which is caused by the non-linear voltage of the rf system.
In general, LLD occurs when the frequency of the coherent bunch oscillations emerges from the incoherent band of the bunch synchrotron frequencies. 
Different techniques are employed to enlarge the synchrotron frequency spread, including increasing the longitudinal emittance. However, the most effective method is introducing higher harmonic rf systems~(these are often referred to as Landau rf systems).
Several accelerators, including the Proton Synchrotron~(PS) and Super Proton Synchrotron~(SPS) at CERN, operate with double-harmonic rf systems to increase the instability threshold~\cite{Argyropoulos:2285796,Bohl:360139,spsElena,elenaLHC,Burov1,Frankenheim,PhysRevSTAB.10.104202} or manipulate the bunch shape. 

\begin{figure*}[t]
    \centering
    \includegraphics*[scale=1]{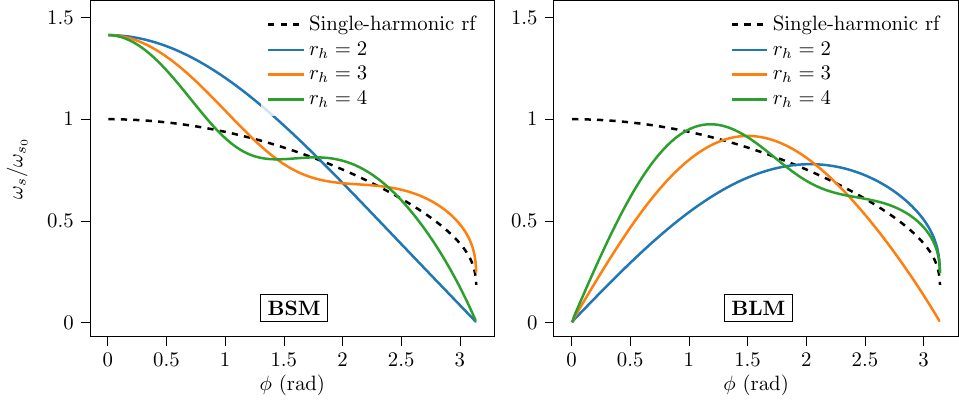}
    \caption{Synchrotron frequency distribution, for different harmonics $r_{h}$, normalized to the small-amplitude synchrotron frequency in single-harmonic rf. Both operating modes (BSM left-hand side instead of BLM right-hand side) are reported as a function of the maximum phase deviation of the particle with $r_{V}=1/r_{h}$. The curves are compared with the conventional single rf system case (black dashed line).}
    \label{fig:synchfreq}
\end{figure*}

In double-harmonic rf, the particles experience a total voltage given by
\begin{equation}
V_{\text{rf}} (\phi)=V_{0} [ \sin (\phi+\phi_{s0})+r_{V} \sin \left(r_h \phi+ r_h \phi_{s0}+\Phi_{2}\right)]
\label{eq:TotVolt} \, ,
\end{equation}
where $V_{0}$, is the voltage magnitude of the main rf system, $\phi_{s0}$ is the phase of the synchronous particle, and $\phi$ is the phase offset relative to the synchronous phase.
The parameters $r_{V}$ and $r_{h}$ denote the rf systems voltage and harmonic number ratios, while $\Phi_2$ represents the relative phase between them.
Two distinct operational modes can be distinguished depending on the value of~$\Phi_2$. Bunch shortening mode~(BSM) occurs when the rf systems are in phase at the bunch position for a non-accelerating bucket, leading to shorter bunches. On the other hand, when the rf systems are in counter-phase, one refers to the bunch lengthening mode~(BLM). Figure~\ref{fig:synchfreq} shows the synchrotron frequency distributions for different harmonic ratios, $r_{h}$, in both configurations.
Several methods have been developed to assess the LLD threshold, one of which is the Sacherer stability diagram derived from a general integral equation~\cite{Sacherer1973}.
In this approach, coherent modes are calculated assuming a linear rf field and neglecting the incoherent frequency spread. The LLD occurs when the coherent mode crosses the boundary of the stability diagram. 
Another method, similar to the Sacherer analytical criterion, obtains the LLD threshold from the integral form of the Lebedev equation~\cite{Lebedev} as shown in~\cite{balbekov1990influence} assuming a constant reactive impedance. 

Hoffman and Pedersen proposed an alternative procedure~\cite{Pedersen} to compute the frequency of a rigid-dipole mode, including synchrotron frequency spread in a single-harmonic rf system and elliptic particle distribution.
Boine-Frankenheim, Shukla, and Chorniy~\cite{Frankenheim,PhysRevSTAB.10.104202} extended the Hofmann-Pedersen approach to the case of multi-harmonic rf system.

The Sacherer stability criterion and the Hoffman-Pedersen approach
can overestimate the actual LLD threshold due to inherent approximations as observed in calculations~\cite{Burov1} and confirmed in measurements~\cite{elenaLHC}.
 
Chin, Satoh, and Yokoya~\cite{Chin1983} contributed to a new description of Landau damping, explaining it as the phase mixing of the van Kampen modes~\cite{vkmode,vkmode2}. The LLD occurs when any of these modes emerge beyond the boundaries of the incoherent synchrotron frequencies band. Burov~\cite{Burov1} applied this concept together with an eigenvalue approach by Oide and Yokoya~\cite{oide1990longitudinal} to study the LLD threshold considering potential well distortion and synchrotron frequency spread.

Recent work exploited the Lebedev equation and emerged van Kampen mode criterion to determine the LLD thresholds in a single-harmonic rf system~\cite{Karpov2021}. Unlike prior studies, it has been observed that for particle distributions belonging to the binomial family, a pure inductive impedance results in a zero LLD threshold unless an upper cutoff to the impedance is introduced.
Furthermore, both analytical and numerical studies have indicated that above transition energy (or capacitive impedance below), the LLD threshold becomes inversely proportional to the cutoff frequency, $f_c$ when $f_c \gg  1/\tau_{\text{full}}$ ($\tau_{\text{full}}$ represents the full bunch length).

Section~\ref{sec2} outlines the fundamental equations governing longitudinal motion and provides solutions for stationary particle distribution. This section also introduces the perturbation formalism, from which the Lebedev matrix equation and the Oide-Yokoya method are derived.

Section~\ref{sec3} provides a concise overview of the formalism related to emerged van Kampen modes. Building upon the work~\cite{Karpov2021}, we present an analytical expression for the LLD threshold in the presence of inductive impedance, derived from the Lebedev equation within the context of double-harmonic rf systems in BSM. The equation is then compared with numerically determined LLD thresholds obtained by solving the matrix equations using the Oide-Yokoya method and, subsequently, tested with macroparticle tracking simulations. Examples of the Large Hadron Collider (LHC) parameters are provided to enable a direct comparison with the findings in~\cite{Karpov2021}. The section continues analyzing the LLD threshold in BLM, highlighting the main limitation arising from the presence of a maximum in the synchrotron frequency distribution. Eventually, the main impact on the threshold due to phase misalignment between the rf systems will be discussed.
The work progresses in Sec.~\ref{sec:brk}, where the evaluation of the beam response to a rigid-dipole perturbation is presented as an observable for beam-based measurements to estimate the LLD threshold. This method was subsequently applied in two distinct accelerators operating under different beam conditions, SPS and PS. The findings of this extensive measurements campaign are compared with semi-analytical calculations using the MELODY code~\cite{MELODY} and macroparticle tracking simulations in BLonD~\cite{BLonD}.

\section{FUNDAMENTAL EQUATIONS OF MOTION AND DEFINITIONS} \label{sec2}
This section introduces the main equations and definitions utilized in this work. After introducing the equations of motion and the adopted coordinate system, we establish the criteria for the perturbed dynamics, which will form the basis for the LLD derivation discussed in the subsequent section.

\subsection{Steady state}\label{subsec:s-s} 
For convenience, this paper adopts the coordinate system~$(\mathcal{E}, \psi)$ involving the energy and phase of the synchrotron oscillation to describe longitudinal beam motion
\begin{equation}
\begin{gathered}
\mathcal{E}=\frac{\dot{\phi}^{2}}{2 \omega_{s0}^{2}}+U_{t}(\phi) \\
\psi=\operatorname{sgn}(\eta \Delta E) \frac{\omega_{s}(\mathcal{E})}{\sqrt{2} \omega_{s0}} \int_{\phi_{\max }}^{\phi} \frac{d \phi^{\prime}}{\sqrt{\mathcal{E}-U_{t}\left(\phi^{\prime}\right)}}
\end{gathered}
\label{eq:bld} \, ,
\end{equation}
where $\Delta E$ and $\omega_{s0}$ are the energy offset with respect to the synchronous particle and the angular frequency of small amplitude synchrotron oscillations in the single harmonic rf system. The slip factor is represented by $\eta= 1/\gamma_{\text{tr}}^{2}-  1/\gamma^{2}$ in which  $\gamma_{\text{tr}}$ is the Lorentz factor at transition energy.
The total potential $U_{t}$, in Eq.~\eqref{eq:bld}, is related to the net voltage according to
\begin{equation}
U_{t}(\phi)=\frac{1}{V_{0} \cos \phi_{s 0}} \int_{\Delta \phi_{s}}^{\phi} \left[V_{t}\left(\phi^{\prime}\right)-V_{0} \sin \phi_{s 0}\right]d \phi^{\prime} \, ,
\end{equation}
with the assumption that the synchronous phase shift $\Delta \phi_{s}$, due to intensity effects, must fulfill the relation $V_0 \sin(\phi_{s0})= V_\mathrm{rf} (\phi_{s0}+\Delta \phi_{s})+V_{\text{ind}}(\Delta \phi_{s})$, where $V_{\text{ind}}(\Delta \phi_{s})$ is the induced voltage.

The synchrotron frequency $\omega_{s} (\mathcal{E})$ can be found by means of the synchrotron period $T_{s}(\mathcal{E})$, according to the expression
\begin{equation}
T_{s}(\mathcal{E})=\frac{2 \pi}{\omega_s(\mathcal{E})}=\frac{\sqrt{2}}{\omega_{s 0}} \int_{\phi_{\min }(\mathcal{E})}^{\phi_{\max }(\mathcal{E})} \frac{d \phi^{\prime}}{\sqrt{\mathcal{E}-U_{t}\left(\phi^{\prime}\right)}}
\label{eq:period} \, ,
\end{equation}
where $\phi_{\min}(\mathcal{E})$ and $\phi_{\max}(\mathcal{E})$ denote the minimum and maximum phase reachable by a particle during its oscillation.

In general, the line density is defined by the distribution function $\mathcal{F}(\mathcal{E})$, namely

\begin{equation}
\lambda(\phi)=2 \omega_{s 0} \int_{U_{t}(\phi)}^{\mathcal{E}_{\max }} \frac{\mathcal{F}(\mathcal{E})}{\sqrt{2\left[\mathcal{E}-U_{t}(\phi)\right]}} d \mathcal{E}
\label{eq:lambda} \, ,
\end{equation}
in which the normalization $\int_{-\pi h}^{\pi h} \lambda(\phi)  d \phi=1$  is imposed. Hereafter, beams with binomial particle distributions will be mostly considered covering most of the realistic longitudinal bunch distributions in proton synchrotrons (e.g. from flat bunches, i.e. $\mu=-1/2$, to Gaussian shape for $\mu 	\rightarrow \infty$)
\begin{equation}
    \mathcal{F}(\mathcal{E})=\frac{1}{2 \pi \omega_{s 0} A_{N}}\left(1-\frac{\mathcal{E}}{\mathcal{E}_{\max }}\right)^{\mu}=\frac{g(\mathcal{E})}{2 \pi \omega_{s 0} A_{N}}
    \label{eq:binomDistr} \, ,
\end{equation}
with normalization factor
\begin{equation}
A_{N}=\omega_{s 0} \int_{0}^{\mathcal{E}_{\max }} \frac{g(\mathcal{E})}{\omega_{s}(\mathcal{E})}d \mathcal{E}
\label{eq:normFactor} \, .
\end{equation}

The integral in Eq.~\eqref{eq:lambda} can be now computed analytically using Eq.~\eqref{eq:binomDistr}, leading to

\begin{equation}
\lambda(\phi)=\frac{\sqrt{\mathcal{E}_{\max }} \Gamma(\mu+1)}{\sqrt{2 \pi} A_{N} \Gamma(\mu+3 / 2)}\left[1-\frac{U_{t}(\phi)}{\mathcal{E}_{\max }}\right]^{\mu+1 / 2}
\label{eq:lambdabinomial} \, ,
\end{equation}
where $ \Gamma$ is the gamma function. For finite $\mu$, the corresponding full bunch length can  be easily derived as
\begin{equation}
\tau_{\text {full }}~=~\left[\phi_{\max }\left(\mathcal{E}_{\max }\right)-\phi_{\min }\left(\mathcal{E}_{\max }\right)\right] / \omega_{\mathrm{rf}} \, ,
\end{equation}
in which $\omega_{\text{rf}}$ represents the angular frequency of the main rf system.

\subsection{Perturbation criteria}
In order to study beam stability, it is essential to account for perturbations relative to the stationary solution discussed in Sec.~\ref{subsec:s-s}. Hence, we introduce $\tilde{\mathcal{F}}$,$\tilde{\lambda}$, and $\tilde{U}_{\text {ind }}$ representing the perturbations of the equilibrium distribution function, line density, and induced potential. The most common approach to studying beam stability is to solve the linearized Vlasov equation, which, in the coordinate system $(\mathcal{E}, \psi)$, takes the following form

\begin{equation}
\left[\frac{\partial}{\partial t}+\omega_{s} \frac{\partial}{\partial \psi}\right] \tilde{\mathcal{F}}=\omega_{s} \frac{\partial \tilde{U}_{\text {ind }}}{\partial \psi} \frac{d \mathcal{F}}{d \mathcal{E}}
\label{eq:vlasov} \, .
\end{equation}

Lebedev proposed \cite{Lebedev} the first self-consistent system of equations suitable for the eigenvalue analysis of longitudinal beam stability. In the following subsection, we present its matrix equation which will allow us to calculate LLD thresholds for single bunches.

\subsubsection{Lebedev equation}

A detailed derivation of the Lebedev equation is documented in~\cite{Shaposhnikova, Karpov2021}.
It appears as an infinite system of equations for harmonics of the line density perturbation $\tilde{\lambda}_{p}$  at frequency  $\Omega$, i.e.
\begin{equation}
\tilde{\lambda}_{p}(\Omega)=\frac{\zeta}{h \cos{\phi_{s0}}} \sum_{k=-\infty}^{\infty} G_{p k}(\Omega) Z_{k}(\Omega) \tilde{\lambda}_{k}(\Omega)
\label{eq:lebedev} \, .
\end{equation}
In the context of circular accelerators, the harmonic of a beam-coupling impedance $Z$ is $Z_k (\Omega)= Z(k \omega_0 + \Omega)$, and the line density perturbation can then be expressed as~\cite{Karpov2021}
\begin{equation}
\tilde{\lambda}_k(\Omega)=\frac{1}{2 \pi h} \int_{-\pi h}^{\pi h}  \tilde{\lambda}(\phi, \Omega) e^{-i k \phi / h} e^{-i \Omega \phi / \omega_{\mathrm{rf}}}d \phi \, .
\end{equation}
Furthermore, the intensity parameter, $\zeta$, in Eq.~\eqref{eq:lebedev}, by definition is equal to
\begin{equation}
\zeta =\frac{q N_{p} h^{2} \omega_{0}}{V_{0} }
\label{eq:zeta} \, ,
\end{equation}
which is proportional to the number of particles $N_p$ and to the square of the fundamental harmonic number in the double-harmonic RF system, $h$. Assuming a binomial particle distribution, as in Eq.~\eqref{eq:binomDistr}, the beam transfer matrices $G_{p k}(\Omega)$~\cite{Shaposhnikova} in Eq.~\eqref{eq:lebedev}, are represented as
\begin{equation}
\begin{aligned}
G_{p k}&(\Omega)=\\
&-i \frac{\omega_{s 0}}{\pi A_{N}} \sum_{m=1}^{\infty} \int_{0}^{\mathcal{E}_{\max }} \frac{d g(\mathcal{E})}{d \mathcal{E}} \frac{I_{m k}^{*}(\mathcal{E}) I_{m p}(\mathcal{E}) \omega_{s}(\mathcal{E})}{\Omega^{2} / m^{2}-\omega_{s}^{2}(\mathcal{E})}d \mathcal{E}
\label{eq:Gpk} \, ,
\end{aligned}
\end{equation}
where $p$ and $k$ denote the revolution frequency harmonics, while $m$ refers to the order of the azimuthal mode (e.g. $m=1$ for the dipole mode; $m=2$ quadrupole mode, etc.). The function $I_{m k}$, introduced in \cite{Lebedev}, can be expressed as
\begin{equation}
 I_{mk}(\mathcal{E}) =  \frac{1}{2 \pi} \int_{- \pi} ^{\pi} e^{i \phi(\mathcal{E}, \psi) k/h + i m \psi} d \psi
\label{eq:Imk} \, .
\end{equation}
Note that the above function is intensity dependent as it is related to the potential well distortion via $\phi(\mathcal{E}, \psi)$.

For a non-trivial solution of the Lebedev equation, the determinant to the associated matrix with the system in Eq.~\eqref{eq:lebedev} must be zero, i.e.

\begin{equation}
D(\Omega, \zeta)=\operatorname{det}\left|\delta_{p k}-\frac{\zeta}{h \cos{\phi_{s0}}} G_{p k}(\Omega) Z_{k}(\Omega) / k\right|=0
\label{eq:determinant} \, .
\end{equation}
Hence, the solution of the above determinant depends on the intensity and specific frequencies $\Omega$.

\subsubsection{Oide-Yokoya method}
A complementary approach was proposed by Oide and Yokoya~\cite{oide1990longitudinal} and has already been applied in beam stability studies as in~\cite{Karpov2021, Burov1, Burov2} allowing to derive the longitudinal bunch oscillation modes. In particular, following the Oide-Yokoya method, the perturbed density function is expanded in Fourier series over the phase:
\begin{equation}
\begin{aligned}
 \tilde{\mathcal{F}}(\mathcal{E},\psi,t)&= \\ 
 & e^{i \Omega t} \sum_{m=1}^{\infty}\left[C_{m}(\mathcal{E}, \Omega) \cos m \psi+S_{m}(\mathcal{E}, \Omega) \sin m \psi\right] \, .
 \end{aligned}
 \label{eq:F_expansion}
\end{equation}
The functions $C_{m}(\mathcal{E}, \Omega)$ are discretized~\cite{oide1990longitudinal} by multiplying them with step-like functions $s_n(\mathcal{E})$, where $\mathcal{E}_n$ is the $n$-th sample of the energy grid, i.e.
\begin{equation}
    C_m (\mathcal{E}, \omega)= \sum_{n=0}^{N_{\mathcal{E}}} s_n(\mathcal{E})C_m (\mathcal{E}_n, \omega) \, ,
    \label{eq:discretizaion}
\end{equation}
where
\begin{equation}
s_n(\mathcal{E})= \begin{cases}1 / \Delta \mathcal{E}_n, & \mathcal{E}_n-\Delta \mathcal{E}_n / 2<\mathcal{E} \leq \mathcal{E}_n+\Delta \mathcal{E}_n / 2 \\ 0, & \text { elsewhere }\end{cases} \, ,
\end{equation}
and $\Delta \mathcal{E}_n$ is the thickness of the corresponding step.
However, the MELODY code~\cite{MELODY} implements a refined non-uniform mesh, with a total number of points $N_{\mathcal{E}}$, to have a high-resolution matrix in energy and frequency when close to the center or edge of the total potential.
Inserting the Eqs.~\eqref{eq:F_expansion} and~\eqref{eq:discretizaion} into the Vlasov equation~\eqref{eq:vlasov}, yields~\cite{oide1990longitudinal}
\begin{equation}
\Omega^{2} C_{m}\left(\mathcal{E}_{n}, \Omega\right)=\sum_{n^{\prime}=1}^{N_{\mathcal{E}}} \sum_{m^{\prime}=1}^{m_{\max }} M_{n m n^{\prime} m^{\prime}} C_{m^{\prime}}\left(\mathcal{E}_{n^{\prime}}, \Omega\right)
\label{eq:oymethod} \, .
\end{equation}
Equation~\eqref{eq:oymethod} represents an eigenvalue problem of linear algebra in which the matrix $M_{n m n^{\prime} m^{\prime}}$ is
\begin{equation}
\begin{aligned}
M_{n m n^{\prime} m^{\prime}} = \; &m^{2} \omega_{s}^{2}\left(\mathcal{E}_{n}\right) \delta_{n n^{\prime}} \delta_{m m^{\prime}} \\
&-\frac{2 \zeta m^{2} \omega_{s}^{2}\left(\mathcal{E}_{n}\right) \omega_{s 0} \Delta \mathcal{E}_{n^{\prime}}}{\pi A_{N} \omega_{s}\left(\mathcal{E}_{n^{\prime}}\right)} \frac{d g}{d \mathcal{E}}\left(\mathcal{E}_{n}\right) \\
& \times \operatorname{Im}\left\{\sum_{k=1}^{k_{\max }} \frac{Z_{k} / k}{h} I_{m k}^{*}\left(\mathcal{E}_{n}\right) I_{m^{\prime} k}\left(\mathcal{E}_{n^{\prime}}\right)\right\}
\end{aligned}
\label{eq:matrixOY} \, ,
\end{equation}
where $\delta_{n n^{\prime}}$ is the Kronecker delta while $m_{\max}$ and $k_{\max}$ are the maximum values of the azimuthal mode and the revolution harmonic number, respectively. In general, the exact solution requires infinite numbers of azimuthal modes and harmonic revolution numbers. However, for numerical reasons, a truncation must be taken into account.

\section{Loss of Landau Damping in Double-Harmonic rf systems} \label{sec3}
In this section, we will study the LLD threshold for the most common rf configurations of double-harmonic rf systems (i.e., in BSM and BLM). Inductive impedances often dominate hadron synchrotron accelerators (e.g., PS and SPS). Therefore, for the sake of simplicity, a pure inductive impedance above transition energy $\eta\text{Im}Z/k>~0$ has been chosen. Nonetheless, expanding the study to a more realistic case may be possible by introducing the concept of effective impedance~\cite{Karpov2021,NeseEffectiveImp, Karpov2022wso}.

For the BSM, an analytical equation of the LLD threshold will be derived by solving the Lebedev equation and, subsequently, compared with a semi-analytical approach using the MELODY code, as well as with macroparticle simulation in BLonD~\cite{BLonD}.

Eventually, the impact on the LLD threshold of phase errors between the rf system will be discussed.

\subsection{The LLD threshold  for pure inductive impedance in BSM}
Following the van Kampen modes description, an eigensystem of the Vlasov equation was found in an infinite plasma revealing a spectrum composed of continuous and discrete parts~\cite{vkmode, vkmode2}. The continuous spectrum is characterized by singular functions related to single-particle motion, whereas the discrete spectrum may not necessarily exist.

In the context of beam dynamics, the concept of van Kampen modes is used to describe bunch oscillations in the longitudinal plane~\cite{Chin1983}. In particular, it has been shown that at low intensity (i.e., $\zeta\approx 0$), the complete set of van Kampen modes is entirely contained within the continuous spectrum, $\Omega=m \omega_s(\mathcal{E})$. In this case, Landau damping results from the decoherence of these modes, which do not represent the collective dynamic but rather the single-particle motions. However, as the intensity increases, Landau damping can be lost and the discrete van Kampen modes emerge from the incoherent band lying beyond $\omega_s(\mathcal{E})$. 

\subsubsection{Analytical criteria}

Following the procedure in~\cite{Karpov2021}, we present an approximate analytical equation of the LLD threshold in BSM for the case of $\eta \text{Im}Z/k>0$ (this covers most of the common scenarios for high-energy synchrotrons e.g., PS and SPS). For simplicity, the threshold is derived by solving the Lebedev equation~\eqref{eq:lebedev} for $\text{Im}Z/k=\text{const}$.

At low intensities, the synchrotron frequency distribution in BSM is a monotonic function of the synchrotron oscillation energy, $\mathcal{E}$. Assuming this holds at the LLD threshold for the fundamental azimuthal mode $m=1$~(dipole mode), we can determine the intensity  $\zeta_{\text{th}}$ at which the coherent mode frequency $\Omega$ equals the maximum incoherent frequency. This entails solving Eq.~\eqref{eq:determinant} to find when ${{\Omega=\max\left[\omega_{s}(\mathcal{E})\right]}=\omega_{s}(0)}$.

According to linear algebra, the determinant of the matrix in Eq.~\eqref{eq:determinant} can be expressed as
\begin{equation}
\operatorname{det}(I-\zeta X)= \prod_{n=0} (\nu_n -1) = 0
\label{eq:Matrix_det} \, ,
\end{equation}
where $I$ is the identity matrix
and $\nu_n$ is the $n$th eigenvalue of matrix $\zeta X$. Following Eq.~\eqref{eq:Matrix_det}, one of the eigenvalues has to equal one. Hence, assuming, e.g., $\nu_0=1$ and based on the property of the matrices trace $\operatorname{tr}(\zeta X) = \sum_n \nu_n $, we obtain
\begin{equation}
    \zeta = \frac{1+ \varepsilon}{\operatorname{tr}(X)} \, ,
    \label{eq:zeta_trace}
\end{equation}
where $ \varepsilon=\sum_{n\neq0}\nu_n $ and 
\begin{equation}
    \operatorname{tr}(X)= \frac{1}{h\cos{\phi_{s0}}} \sum_{k=-\infty}^{\infty} G_{k k}(\Omega) \frac{Z_{k}(\Omega)}{k} \, .
    \label{eq:traceX}
\end{equation}
Typically, in a binomial distribution, the assumption, $\varepsilon \ll 1 $ is valid \cite{PhysRevAccelBeams.27.074401}. This allows us to reduce Eq.~\eqref{eq:zeta_trace} as
\begin{equation}
    \zeta_{\mathrm{th}} \approx 1/\operatorname{tr}(X) \, .
    \label{eq:approx_zeta}
\end{equation}
Substituting Eq.~\eqref{eq:traceX} into Eq.~\eqref{eq:approx_zeta} leads to a general approximate LLD threshold for binomial particle distributions
\begin{equation}
\zeta_{\mathrm{th}}=h \cos{\phi_{s0}}\left[\sum_{k=-\infty}^{\infty} G_{k k}(\Omega) \frac{Z_{k}(\Omega)}{k}\right]^{-1}
\label{eq:zetath1} \, .
\end{equation}

The synchrotron frequency distribution can generally be derived from Eq.~\eqref{eq:period}. However, finding an exact closed solution for a double-harmonic rf configuration is not trivial, and numerical solutions are often calculated instead.
Following the work in~\cite{ActiveL}, an analytical equation for the synchrotron frequency distribution, as a function of synchrotron oscillation energy, is derived in BSM for small amplitude oscillation ($\mathcal{E} \ll 1$). We obtain
\begin{equation}
\frac{\omega_{s}(\mathcal{E})}{\omega_{s0}}=\sqrt{1+r_{V} r_{h} }\left(1-\frac{1+r_{V} r_{h}^{3}}{(1+r_{V} r_{h})^{2}} \frac{\mathcal{E}}{8}\right)
\label{eq:feqDRF} \, .
\end{equation}
Furthermore, thanks to the low energy oscillation,  ${\phi(\mathcal{E},\psi)\approx \sqrt{2 \mathcal{E}/(1 + r_{V} r_{h})} \cos{\psi}}$. This allows us to simplify Eq.~\eqref{eq:Imk} into a Bessel function of the first kind, i.e.
\begin{equation}
    I_{mk}\approx i^m J_m \left(\frac{k}{h}\sqrt{\frac{2 \mathcal{E}}{1 + r_{V} r_{h}}}\right)
    \label{eq:Imk_bessel} \, .
\end{equation}

Based on Eq.~\eqref{eq:feqDRF}, the LLD occurs when ${\Omega=\omega_{s0}\sqrt{1+r_{V} r_{h}}}$. We can compute the matrices $G_{kk}$ keeping only the first element of the sum over the azimuthal modes ($m=1$), as well as including Eq.~\eqref{eq:Imk_bessel}. Thus, the Eq.~\eqref{eq:Gpk} becomes

\begin{equation}
\begin{aligned}
G_{k k}=&-\frac{8 i \sqrt{1 + r_{V} r_{h}}}{\pi A_{N} \phi_{\max }^{2} (1+r_{V} r_{h}^{3})} \int_{0}^{1}  \frac{d g(x)}{d x} J_{1}^{2}\left(\frac{k x}{h} \phi_{\max}\right) \\
& \times \frac{(1+r_{V} r_{h})/(1+r_{V} r_{h}^3)-\phi_{\max }^{2} x^{2} / 16   }{x^{2} (1+r_{V} r_{h})/(1+ r_{V} r_{h}^3) -\phi_{\max }^{2} x^{4} / 32} d x
\end{aligned} \label{eq:Gkk_dipole} \, .
\end{equation}
Note that a change of variable was performed by imposing ${x=\sqrt{\mathcal{E}/\mathcal{E}_{\max}}}$; moreover, we defined the maximum phase amplitude as ${\phi_{\max}=\sqrt{2 \mathcal{E}_{\max}/(1+r_{V} r_{h})}}$. According to Eq.~\eqref{eq:normFactor} and employing Eq.~\eqref{eq:binomDistr}, we can determine the normalization factor for a binomial distribution by neglecting the synchrotron frequency spread, i.e.~${A_N=\sqrt{1+r_{V} r_{h}} ~\phi_{\max}^2 / 2(\mu + 1)}$. Thus, Eq.~\eqref{eq:Gkk_dipole} can be written as

\begin{equation}
G_{k k} \approx i \frac{16 \mu(\mu+1)}{\pi \phi_{\max }^{4} (1+r_{V} r_{h}^{3})}\left[1-{ }_{1} F_{2}\left(\frac{1}{2} ; 2, \mu ;-y^{2}\right)\right]
\label{eq:Gkk_approx} \, ,
\end{equation}
where ${ }_{p} F_{q}\left(a_{1}, \ldots, a_{p} ; b_{1}, \ldots, b_{q} ; z\right)$ is the generalized hypergeometric function and ${y=k \phi_{\max} /h}$. Note that, in Eq.~\eqref{eq:Gkk_approx}, for particular values of $\mu$, the quantity in the square brackets may be expanded as a combination of Bessel functions of the first kind (e.g., $\mu=1$ yields $G_{kk} \propto [1- J_0 ^2 (y)-J_1 ^2 (y)]$).

\begin{figure*}[!htb]
    \centering
  \includegraphics[scale=1]{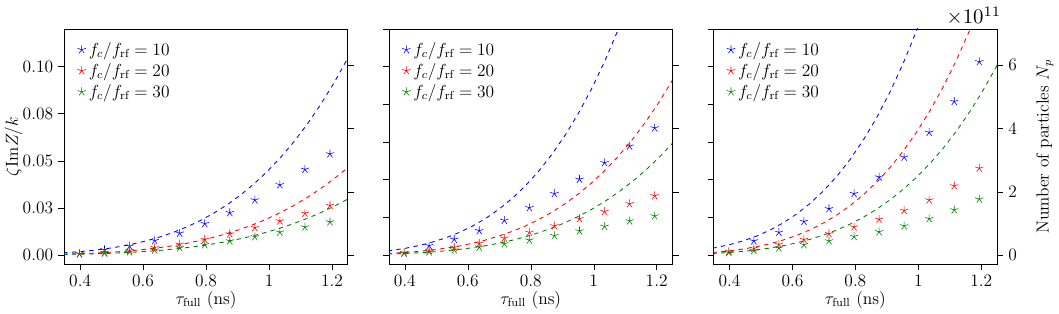}
    \caption{The LLD threshold, for the $2^{\text{nd}}$ (left; $r_{h}=2$), $3^{\text{rd}}$ (center; $r_{h}=3$) and for the $4^{\text{th}}$ (right; $r_{h}=4$) harmonic rf system case, as a function of the full bunch length. Using MELODY for different impedance cutoff frequencies, the thresholds have been calculated and compared with the analytical predictions represented as dashed lines. The computation was performed based on $\text{Im}Z/k=0.07$~$\Omega$, $r_{V}=1/r_{h}$ and $\mu=1.5$.}
    \label{fig:LLD_EQ}
\end{figure*}

Eventually, Eq.~\eqref{eq:zetath1} can be evaluated analytically by approximating the sum as an integral. Therefore, considering a constant inductive impedance, it can be approximated according to
\begin{equation}
\frac{1}{h} \sum_{k=-\infty}^{\infty} G_{k k}(\Omega) \frac{Z_{k}(\Omega) }{k} \approx \frac{i}{h}\text{Im} Z  /k \int_{-\infty}^{\infty} G_{k k}(\Omega)  d k \rightarrow \infty \, ,
\end{equation}
which diverges for $\mu>0$. For finite threshold, a truncation  of the sum at an arbitrary $k_{\max}$ must be taken into account, yielding the final expression
\begin{equation}
    \zeta_{\text{th}} \approx \frac{1+r_{V} r_{h}^{3}}{\mu(\mu+1)}\frac{ \pi \phi_{\max}^{5} }{32 \chi\left(k_{\max}\phi_{\max}/h, \mu\right) \text{Im} Z/k}
    \label{eq:LLDth} \, ,
\end{equation}
where function $\chi\left(y, \mu\right)$, is related to the generalized hypergeometric function, according to
\begin{equation}
\chi(y, \mu)=y\left[1-{ }_{2} F_{3}\left(\frac{1}{2}, \frac{1}{2} ; \frac{3}{2}, 2, \mu ;-y^{2}\right)\right].
\label{eq:chi}
\end{equation}

\noindent Equation~\eqref{eq:LLDth} shows the significant impact of the harmonic number and voltage ratios on the LLD threshold. Removing the second rf system~($r_{V}=0$), the equation reduces to the simplest case of single-harmonic rf~\cite{Karpov2021}. The gain factor, $1+r_V r^3_h$, is the same as the one obtained in the analysis of coupled-bunch instability threshold~\cite{balbekov1987methods}.  In the limit case of $y \rightarrow \infty$, the generalized hypergeometric function converges to zero leading Eq.~\eqref{eq:LLDth} to the following simplified expression
\begin{equation}
    \zeta_{\text{th}} \approx \frac{1+r_{V} r_{h}^{3}}{\mu(\mu+1)} \frac{\pi \phi_{\max}^{4} h}{32 k_{\max}} \frac{1}{\text{Im} Z/k}
    \label{eq:LLDth2} \, .
\end{equation}
Equation~\eqref{eq:LLDth2} highlights the dependency of the threshold on the fourth power of the bunch length and its inverse proportionality on the cutoff frequency.

The closed analytical threshold given by Eq.~\eqref{eq:LLDth} will be compared with a semi-analytical approach employing the MELODY code. The computation is based on finding the eigenvalues of Eq.~\eqref{eq:matrixOY} numerically, for the Oide-Yokoya method, as a function of the intensity parameter, $\zeta$. The threshold will eventually correspond to the intensity at which the maximum incoherent frequency $\hat{\omega}_s$ equals the maximum eigenfrequency $\max(\Omega_{1n})$.

\begin{table}[!hbt]
   \centering
   \caption{Main rf parameters of the LHC~\cite{lhcdesign}.}
   \begin{tabular}{l c r}
       \toprule
       \textbf{Parameter}                     & 
       \textbf{Unit} & \textbf{Value} \\
       \hline
           Circumference, $2 \pi R$ & $\text{m}$                      &  $26658.86 $      \\ %[3pt]
          Main harmonic number, $h $   &               & $35640$        \\ %[3pt]
         Main rf frequency, $f_{\text{rf}}$  & $\text{MHz}$    &    $400.79$  \\
         Beam energy, $E_0$ &$\text{TeV}$ & $0.45$ \\
         Main rf voltage $V_{0} $   & $\text{MV}$&  $6$ \\
         Effective impedance, $\text{Im}Z/k$ &  $\Omega$ & $0.07$ \\
       \toprule
   \end{tabular}
   \label{parametLHC}
\end{table}

\begin{figure}[!htb]
    \centering
\includegraphics[width=\linewidth]{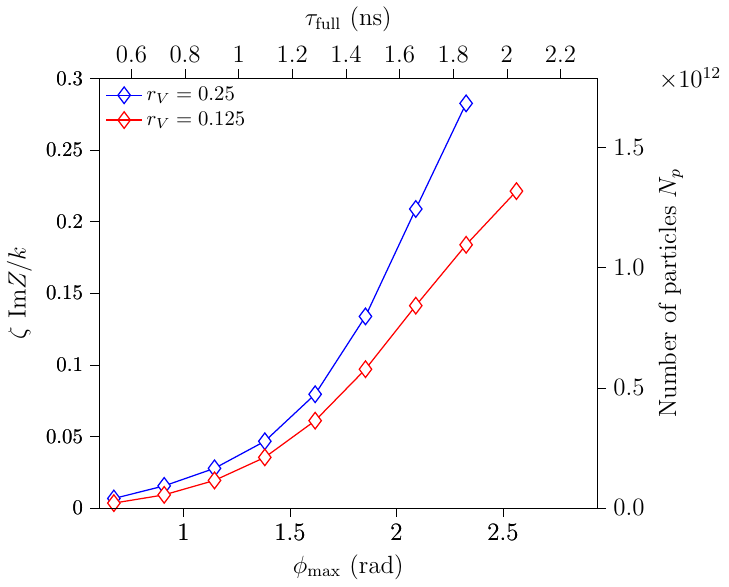}  
    \caption{LLD threshold as a function of the bunch length. The thresholds are computed with MELODY for the voltage ratio $r_{V}=~0.25$ (blue) and $r_{V}=0.125$ (red) by considering Im$Z/k=~0.07~\Omega$ with a cutoff frequency, $f_{c}/f_{\mathrm{rf}}=10$ and binomial coefficient, $\mu=2$.}
    \label{fig:LLD_EQ_lb}
\end{figure}

Although the LHC has no higher-harmonic rf system, in order to facilitate a direct comparison with the analysis conducted in single-harmonic rf~\cite{Karpov2021}, the accelerator parameters outlined in Table~\ref{parametLHC} have been considered.

\begin{figure*}[!htbp]
    \centering
    \includegraphics*[width=\linewidth]{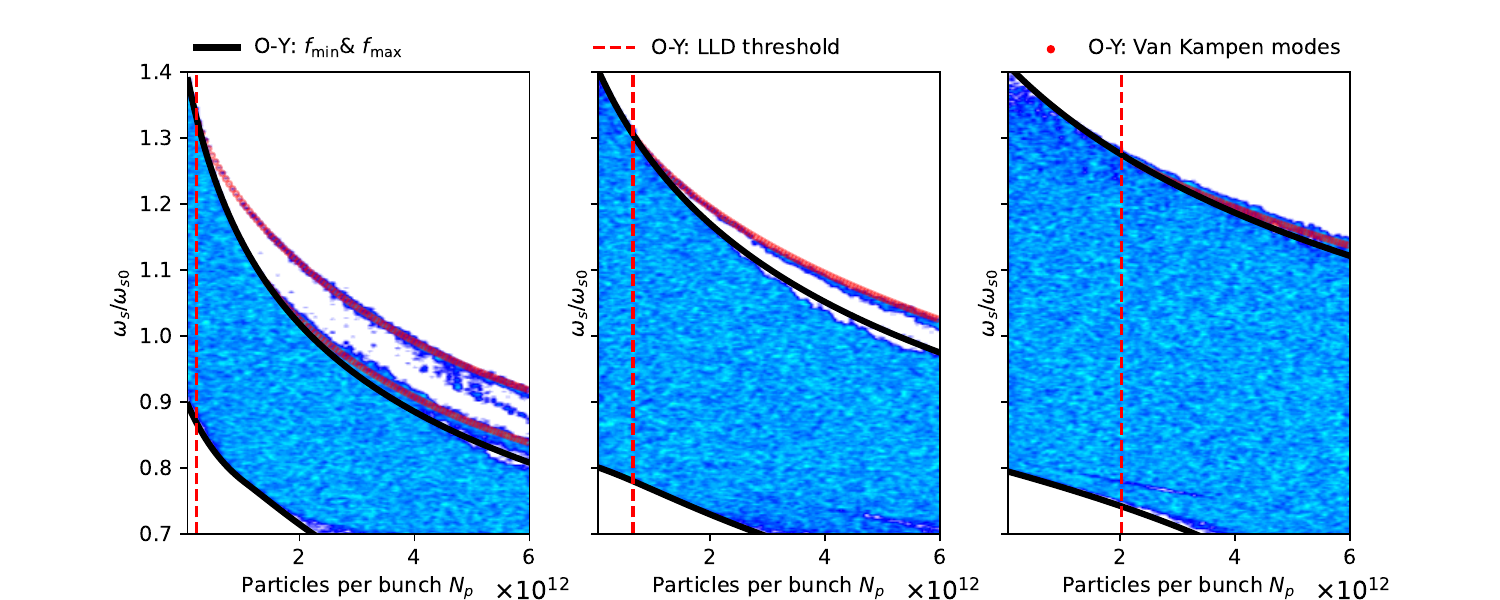}
    \caption{Absolute value of the normalized mode frequency computed in BLonD~(blue color scale) and from MELODY~(red dotted line) as a function of bunch intensity for a constant impedance $\text{Im} Z/k~=~0.07$~$\Omega$ and $f_{c}/f_{\mathrm{rf}}=10$. The minimum and maximum incoherent frequencies are obtained with MELODY and are shown as solid black lines. The dashed red lines indicate the LLD intensity thresholds obtained from calculations performed for three different bunch lengths, $\phi_{\max}=1.0$ $\text{rad}$ (left), $\phi_{\max}=1.5$ $\text{rad}$ (centre) and $\phi_{\max}=2.0$ $\text{rad}$ (right). The rf configuration and bunch distribution are $r_{h}=4$, $r_{V}=0.25$, $\mu = 1.5$ and number of azimuthal modes, $m = 15 $.}
    \label{fig:BLonDSpectrum}
\end{figure*}

Figure~\ref{fig:LLD_EQ} depicts the comparison between the LLD threshold computed using Eq.~\eqref{eq:LLDth} (dashed-line) and MELODY (stars) as a function of the full bunch length for different cutoff frequencies. The comparison has been repeated for three distinct rf configurations: second ($r_{h}=2$; left), third ($r_{h}=3$; center), and fourth ($r_{h}=4$; right) harmonic rf systems. Overall, the proposed analytical equation agrees with the exact solution computed via MELODY and scales inversely with the cutoff frequency, as predicted in Eq.~\eqref{eq:LLDth2}. Nonetheless, some discrepancies arise for large bunch lengths due to the short-bunch approximation utilized in the Lebedev equation in deriving the LLD threshold. Moreover, for $r_{h}=4$ (Fig.~\ref{fig:LLD_EQ}, right) the threshold equation can diverge earlier from the semi-analytical calculation as for the impedance cutoff frequency $f_c/f_{\text{rf}}=10$ (blue, dashed).

Computing the threshold along the full bucket, corresponding to $\phi_{\max}=\pi$,
Fig.~\ref{fig:LLD_EQ_lb} shows that the LLD threshold preserves its monotonic behavior. This does not agree with observations in~\cite{Argyropoulos:2285796}, where the LLD threshold decreases once crossing the zero derivatives of the frequency distribution (e.g., Fig.~\ref{fig:synchfreq}, green). The discrepancy arises from a different definition of the LLD threshold in~\cite{Argyropoulos:2285796}, which is based on the decoherence of the particles following a longitudinal phase kick. A similar approach will be presented in Sec.~\ref{sec:brk} where the beam response to a phase excitation will serve as a measurement technique for the LLD threshold.

\subsubsection{Comparison with simulation}
We moreover compare the results of MELODY with macroparticle tracking simulations in BLonD. 

As the azimuthal modes correspond to specific frequencies arising from the incoherent band of synchrotron frequencies, we study the spectrum of the bunch centroid evolution as an observable for the dipole mode. To perform the simulation in BLonD, a bunch with $10^6$ macroparticles was generated, matched with intensity effects, and then tracked for $10^6$ turns in a double-harmonic rf system with a harmonic ratio of $r_{h}=4$ in BSM and a voltage ratio, $r_{V}=0.25$. The large number of turns ensures a good frequency resolution of the spectrum and covers a sufficient number of synchrotron oscillations. Thereafter, a fast Fourier transform~(FFT)~\cite{nussbaumer1982fast} of the bunch center of mass has been performed.
The comparison was conducted considering a pure inductive impedance of $\text{Im}Z/k=0.07$~$\Omega$ with a cutoff frequency of $f_{c}=4$~GHz, and performing a wide intensity scan to observe the behavior of the coherent mode. 
The tracking process is additionally run for different maximum phase deviations, thereby covering several parts of the synchrotron frequency distribution. In order to better reveal the coherent frequencies in the spectrum, a Hanning filter is applied to the signal. This type of filter is commonly applied in statistics for smoothing~\cite{essenwanger1986elements}.

\begin{figure*}[!tbh]
\begin{subfigure}{.333\textwidth}
 \centering
\includegraphics[width=\linewidth]{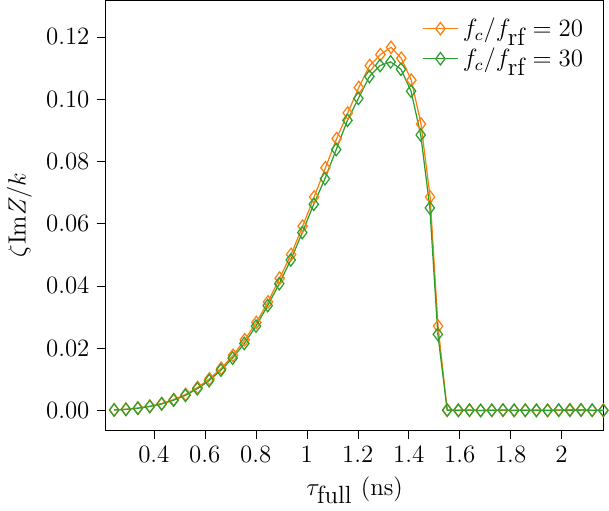}
 \caption{$r_{h}=2; r_{V}=0.4$.}
 \label{fig:BLMn2}
\end{subfigure}%
\begin{subfigure}{.333\textwidth}
  \centering
  \includegraphics[width=\linewidth]{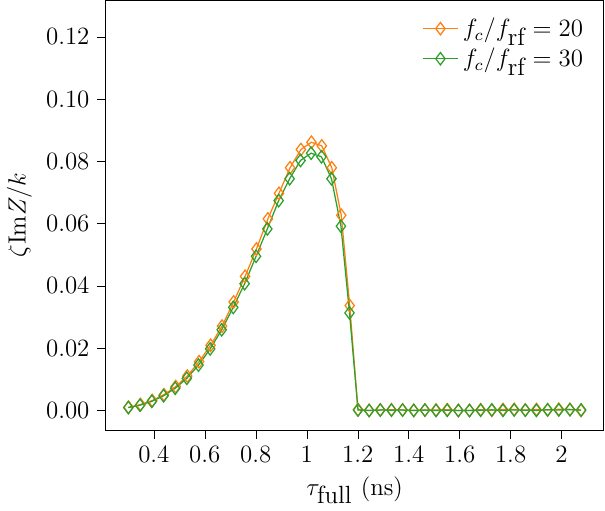}
 \caption{$r_{h}=3; r_{V}=0.3$.}
 \label{fig:BLMn3}
\end{subfigure}%
\begin{subfigure}{.333\textwidth}
  \centering
\includegraphics[width=\linewidth]{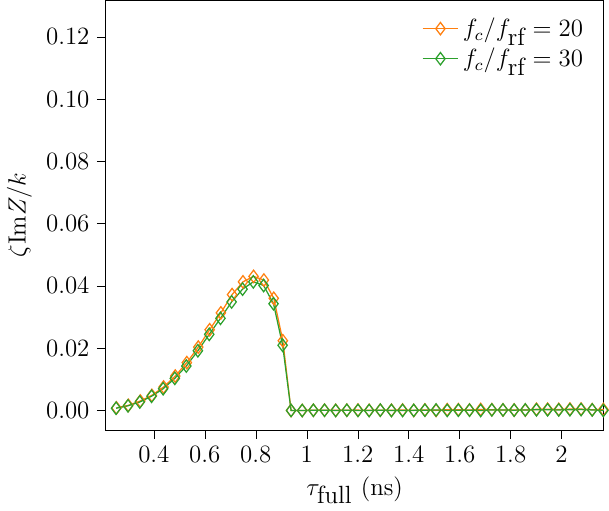}
       \caption{$r_{h}=4; r_{V}=0.2$.}
       \label{fig:BLMn4}
\end{subfigure}%
  \caption{LLD thresholds computed with the MELODY code in BLM for three cases of double-harmonic rf systems: the second harmonic ($r_{h}=2$, left), the third harmonic ($r_{h}=3$, center), and the fourth harmonic ($r_{h}=4$, right) of the fundamental rf system. The thresholds are plotted as a function of the full bunch length, calculated using different impedance cutoff frequencies, with parameters set to $\text{Im}Z/k=0.07$~$\Omega$, and $\mu=1.5$.}
  \label{fig:BLM_LLD}
% \end{subfigure}%
\end{figure*}

Figure~\ref{fig:BLonDSpectrum} shows the normalized mode frequencies for different intensities in BLonD (colors) and the emerging van Kampen modes computed in MELODY (red line) applying the Oide-Yokoya method. Given that the frequency of the mode coincides with the maximum incoherent frequency (black) at the LLD threshold (red, dashed), it becomes impossible in simulation to discriminate the mode properly from the incoherent spectrum. Excellent agreement is shown between BLonD and MELODY, even in the case of a higher-order radial mode of the dipole mode seen as the second emerged mode for $\phi_{\max}=1.0$~rad in Fig.~\ref{fig:BLonDSpectrum}~(left).

\subsection{Threshold of LLD in BLM}

In the context of BLM (rf systems in counter-phase at the bunch position), it becomes non-trivial to determine an analytical closed solution for the LLD threshold since the maximum incoherent frequency, in most cases, varies with the bunch length. Nonetheless, we can still determine the threshold by solving Eq.~\eqref{eq:oymethod} semi-analytically (which has general validity) using the MELODY code.

Contrary to the BSM, Fig.~\ref{fig:BLM_LLD} shows a non-monotonic threshold for a constant inductive impedance, again computed for the reference parameters outlined in Table~\ref{parametLHC}. Previous studies have shown the substantial impact of the zero derivatives of the synchrotron frequency distribution $d{\omega_s}/d{\phi}$~\cite{Burov1, Argyropoulos:2285796}. Once the bunch length exceeds the critical point where $d{\omega_s}/d{\phi}=0$, the LLD vanishes. In BLM, the presence of the discrete mode is significantly influenced by the tails of the distribution.
When the bunch length exceeds the point where $d{\omega_s}/d{\phi}=0$, a van Kampen mode is already located at the maximum of the frequency distribution at zero intensity. Therefore, even for arbitrarily small intensity, the collective effects push the mode outside the incoherent frequency band, causing the LLD.

In conclusion, although BLM provides a larger synchrotron frequency spread (as illustrated in Fig.~\ref{fig:synchfreq}),
we are limited by the presence of zero derivatives in the synchrotron frequency distribution. This constraint on the bunch length progressively becomes more stringent as we increase the order of the harmonic rf system.

\begin{figure}[!tbh]
    \centering
  \includegraphics[width=0.95\linewidth]{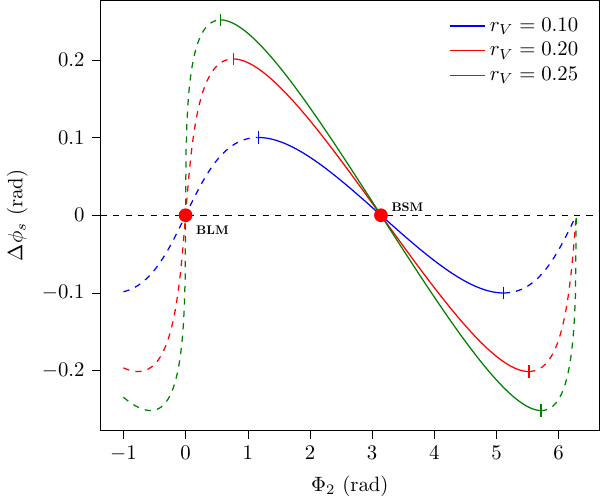}
    \caption{Calculated variation of the synchronous phase shift in a fourth harmonic rf system with respect to relative phase for different voltage ratios, $r_{V}$. The red dots mark the zero-crossing points of the curves, indicating the corresponding working points of BSM and BLM configurations.}
    \label{fig:deltaphis}
\end{figure}

\begin{figure*}[htbp]
\centering
\begin{subfigure}{.5\textwidth}
 \centering
\includegraphics[width=\linewidth]{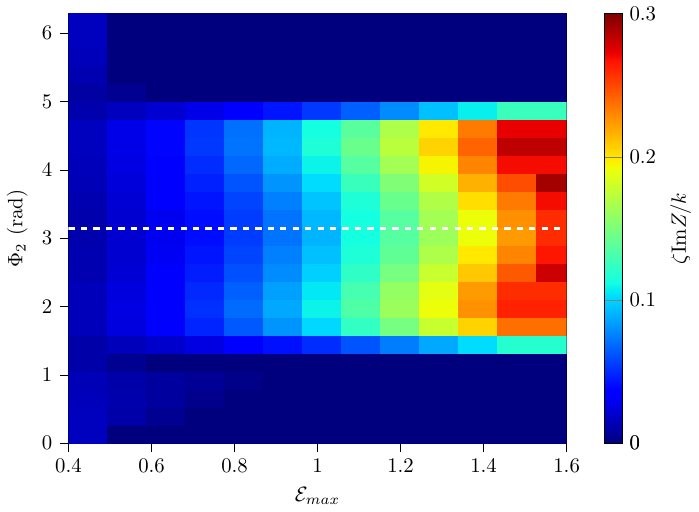}
 \label{fig:PH_r01}
\end{subfigure}%
\begin{subfigure}{.5\textwidth}
  \centering
\includegraphics[width=\linewidth]{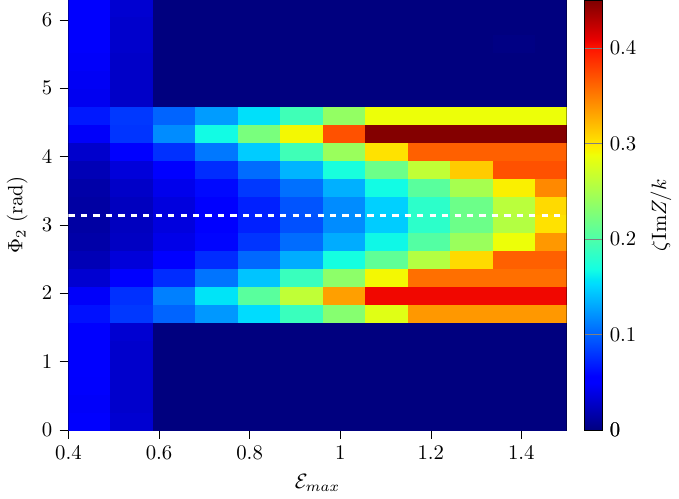}
 \label{fig:PH_r02}
\end{subfigure}%
  \caption{Threshold of LLD (color coding) computed in MELODY as a function of the maximum synchrotron oscillation energy and relative phase, $\Phi_2$. Two representative voltage ratios are considered, namely $r_{V}=0.1$~(left) and $r_{V}=0.2$~(right), with a harmonic number ratio $r_{h}=4$, based on the parameters in Tab.~\ref{parametLHC}. The ideal BSM case is depicted in white.}
  \label{fig:Ph_color}
% \end{subfigure}%
\end{figure*}

\subsection{Relative phase shift} \label{Phase error}

Practically, relative phase errors between  RF systems are difficult to avoid. Therefore, we will examine how the relative phase, $\Phi_2$, in Eq.~\eqref{eq:TotVolt}, impacts the LLD threshold. Figure~\ref{fig:deltaphis} illustrates the computed synchronous phase shift variation, in a fourth harmonic rf system, with respect to the relative phase for various rf voltage ratios, $r_V$. The BSM is less sensitive compared to the BLM where small variations in $\Phi_2$ result in a significant deviation in the synchronous phase and synchrotron frequency. This might represent an additional limitation for operating in the BLM since achieving this accuracy can be extremely challenging.

We computed the LLD threshold along two degrees of freedom in MELODY: relative phase between both rf systems and maximum synchrotron oscillation energy $\mathcal{E}_{\max}$ considering only the fundamental azimuthal mode $m=1$. 
We explored different voltage ratios, namely $r_V=0.1$ (Fig.~\ref{fig:Ph_color}, left), and $r_V=0.2$ (Fig.~\ref{fig:Ph_color}, right), for a harmonic number ratio of $r_h=4$. As expected, the threshold is symmetric along $\Phi_2$, revealing very interesting hybrid configurations with LLD thresholds higher than in the conventional BSM and BLM. Similar to Fig.~\ref{fig:BLM_LLD}, for high synchrotron energy oscillation (e.g. $\mathcal{E}_{\max}>0.6$), moving from $\Phi_2={\pi}$ (white) towards $\Phi_2=0$ and $\Phi_2={2\pi}$, the threshold vanishes.

\section{Beam response from a phase excitation} \label{sec:brk}

\begin{figure*}[!tbh]
\begin{subfigure}{.333\textwidth}
 \centering
\includegraphics[width=\linewidth]{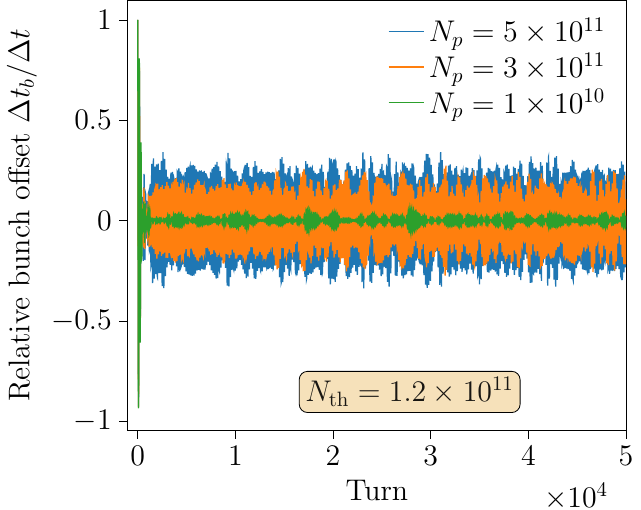}
 % \caption{$n=2; r=0.4$.}
 \label{fig:BKick10}
\end{subfigure}%
\begin{subfigure}{.333\textwidth}
  \centering
\includegraphics[width=\linewidth]{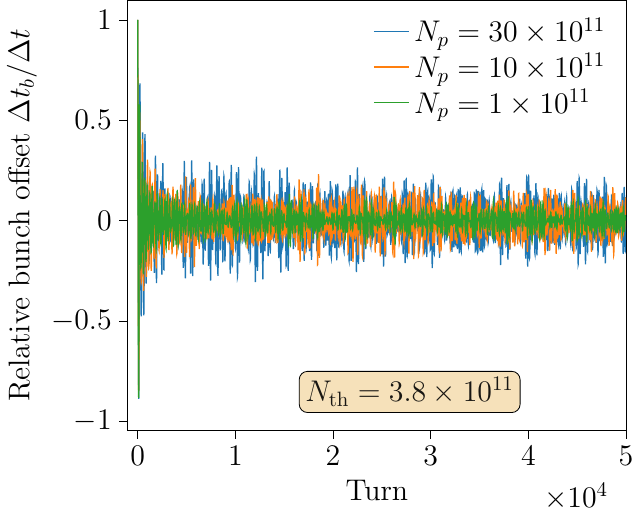}
 % \caption{$n=3; r=0.3$.}
 \label{fig:BKick15}
\end{subfigure}%
\begin{subfigure}{.333\textwidth}
  \centering
\includegraphics[width=\linewidth]{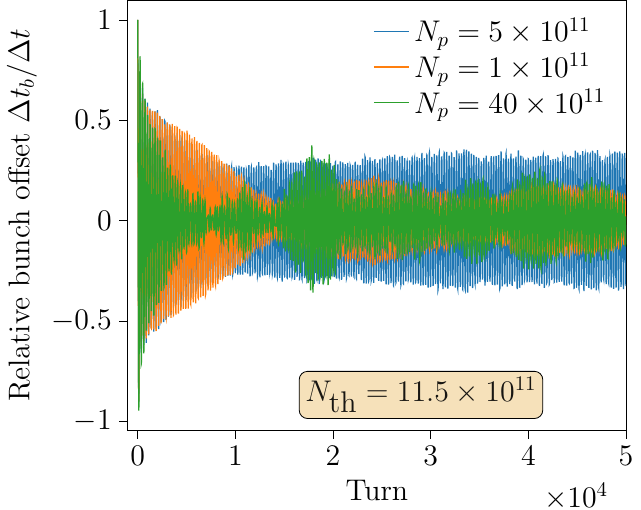}
       % \caption{$n=4; r=0.2$.}
       \label{fig:BKick20}
\end{subfigure}%
  \caption{Simulated bunch offset evolution in BSM, for different intensities, following an initial dipole kick of one degree. The simulation was conducted with BLonD and an inductive impedance of $\text{Im}Z/k=0.07$~$\Omega$ with cutoff at $f_{c}=4$~$\text{GHz}$. Three different bunch lengths, respectively for $\phi_{\max}=1.0$~$\text{rad}$~(left), $\phi_{\max}=1.5$~$\text{rad}$~(centre)  and $\phi_{\max}=2.0$~$\text{rad}$~(right) have been considered. The LLD thresholds predicted by MELODY for all three cases are presented in insets.}
  \label{fig:kick}
\end{figure*}

Investigating the beam response to a longitudinal kick is a commonly employed method for studying LLD through beam measurements and simulations. Furthermore, it also allows probing impedance sources in an accelerator. 

The section introduces the case of a simulated bunch offset evolution for a constant inductive impedance as baseline aspect of the study. This methodology will then be applied to conduct beam-based measurements in the SPS and PS to estimate the LLD threshold at very different conditions. 

\subsection{Simulations for a constant inductive impedance}
To reproduce a rigid-dipole kick, an instant shift of the bunch position by one degree was introduced at the beginning of the macroparticle simulation. Subsequently, the evolution of the bunch distribution was tracked for $5\times10^{4}$ turns. The simulation was conducted using the BlonD code, with the parameters outlined in Table~\ref{parametLHC} and $2\times10^6$ macroparticles.

Figure~\ref{fig:kick} illustrates bunch centroid evolution following an initial phase offset for three different maximum phase deviations (bunch length).
For $\phi_{\max}=1$~rad (left), the initial rapid decoherence of bunch oscillations due to the phase kick is followed by subsequent damping of the bunch centroid amplitude oscillation when the intensity is below the LLD threshold. Above the threshold,  the amplitude of the bunch offset oscillation persists. However, for a relatively large bunch length (such as $\phi_{\max}=2.0$ in Fig.~\ref{fig:kick}, right) the simulations reveal a counter-intuitive behavior that deviates from initial expectations. Specifically, we observe that the damping time can be very long at low intensity (blue). The initial damping becomes faster at higher intensities, but the residual oscillation amplitude remains small (green).

\begin{figure}[!htbp]

    \begin{subfigure}{.445\textwidth}
     \centering
    \includegraphics[width=\linewidth]{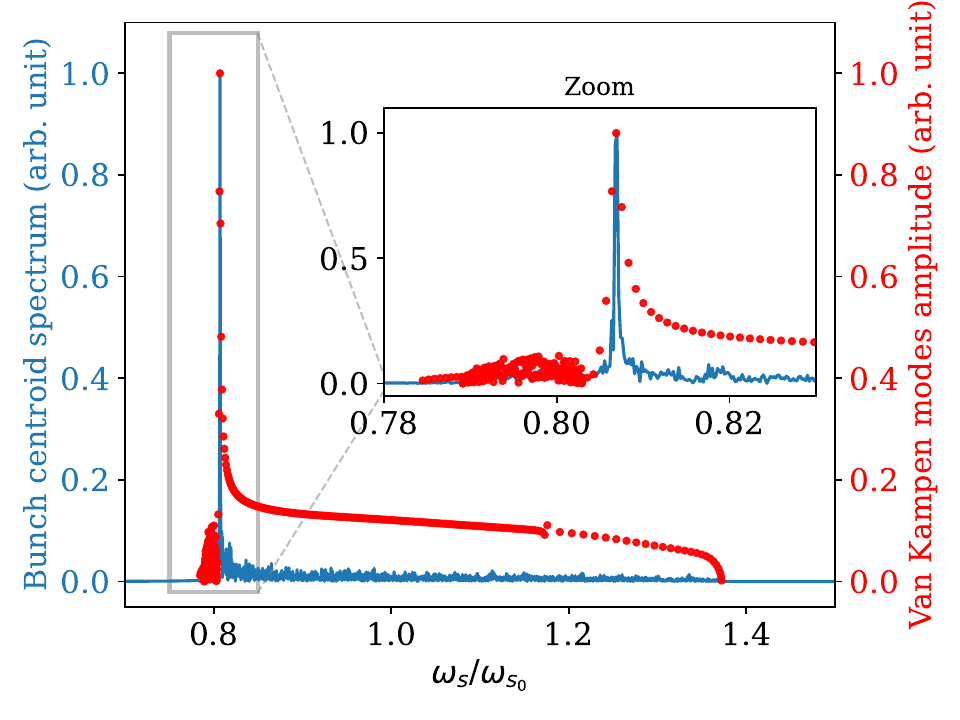}
    \end{subfigure}

\begin{subfigure}{.445\textwidth}
    \centering
    \includegraphics[width=\linewidth]{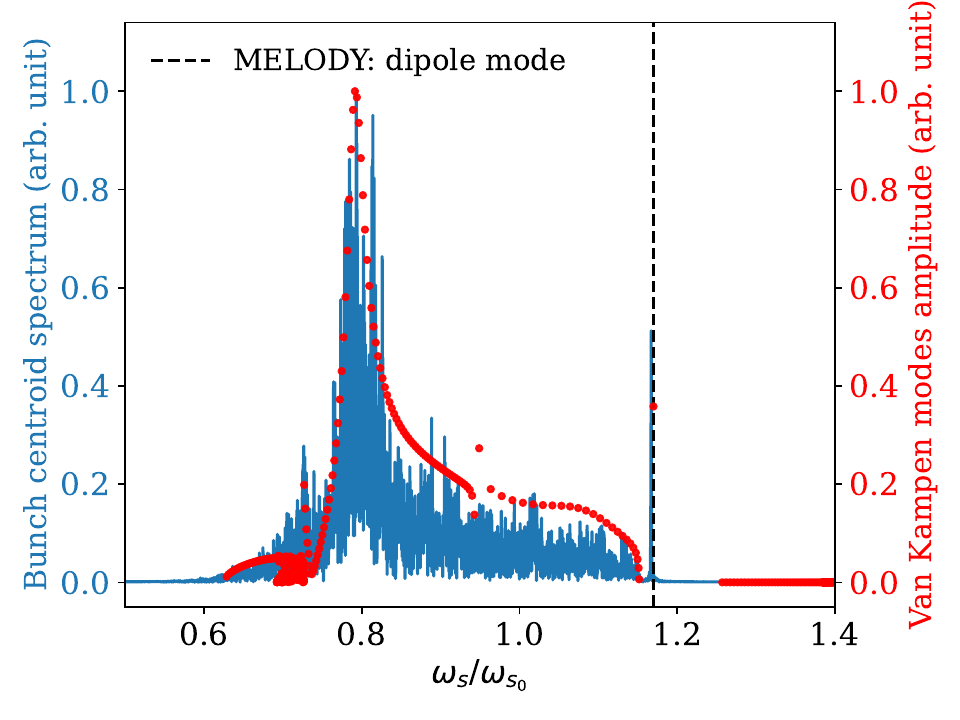}
\end{subfigure}
    \caption{Bunch position spectrum in BLonD (blue) and van Kampen modes in MELODY (red dots). The computations are based on Tab.~\ref{parametLHC} in BSM with $r_{h}=4$, $r_{V}=0.25$, $\phi_{\max}=2$~rad, $\mu=1.5$, and $f_c/f_{\mathrm{rf}}=10$ for the intensities $N_p=5\times 10^{11}$ (top) and $N_p=4\times 10^{12}$ (bottom). The frequency of the dipole mode is highlighted in black.}
    \label{fig:vkm_spectra}
\end{figure}

This behavior can be explained by analyzing the spectrum of the dipolar synchrotron frequency distribution derived from the spectrum of the bunch centroid. Considering the parameters in Tab.~\ref{parametLHC}, the intensity $N_p=4\times 10^{12}$ and maximum phase deviation $\phi_{\max}=2.0$~rad is beyond the flat zone (where $d \omega_s/ d \phi >0$) of the synchrotron frequency distribution (Fig.~\ref{fig:synchfreq}) and also well above the LLD threshold. Similarly to Fig.~\ref{fig:kick}, $2\times 10^{6}$ macroparticles for $10^6$ turns were tracked for a sufficiently high resolution in the frequency domain.
The spectrum of the bunch centroid motion, presented in Fig.~\ref{fig:vkm_spectra}, confirms that LLD is lost for an intensity of $N_p=4\times 10^{12}$ (bottom) since a coherent mode lies outside the incoherent frequency band~(blue). This is also in agreement with the prediction by MELODY (black). Solving Eq.~\eqref{eq:oymethod} with the same parameter used in simulation shows that the van Kampen modes (red dots) behave similarly to the bunch centroid spectrum. The amplitude of the van Kampen modes~\cite{Karpov2021} indicates the strength of the response to a phase kick. Frequencies belonging to the incoherent band with high amplitudes with respect to the coherent mode are shown around $\omega_s/\omega_{s0}\approx0.8$. This coincides with the flat zone of the synchrotron frequency distribution in Fig.~\ref{fig:synchfreq}, where $d \omega_s/ d \phi >0$. It implies that when a perturbation is present, these high-amplitude van Kampen modes within the incoherent spectrum can phase mix. As a result, this reduces the residual amplitude oscillation, as illustrated in Fig.~\ref{fig:kick}. However, for lower intensity, as $N_p=5\times 10^{11}$ in Fig.~\ref{fig:synchfreq}~(top), only a few high-amplitude modes are phase mixing, thereby leading to slower damping.

\subsection{Beam-based measurements}
We will now introduce a simple measurement technique for evaluating LLD in synchrotrons.
The method is applied in the SPS comparing the findings with numerical predictions in BLonD and MELODY. Furthermore, the study is complemented by performing the measurements in the PS with a very different bucket-filling factor. Thanks to the excellent reproducibility and control of rf and beam parameters, this measurement also serves to validate the scaling factor in Eq.~\eqref{eq:LLDth}.

\subsubsection{Longitudinal LLD in the SPS at 200~GeV}
The SPS~\cite{spsElena} is the largest LHC injector at CERN that operates with six $200~\text{MHz}$ traveling wave structures as the main accelerating system and two traveling wave structures at $800~\text{MHz}$ (four times the fundamental rf frequency), supported by several feedback systems against beam instabilities.
Table~\ref{paramet} summarizes the main accelerator parameters used for the measurements.

\begin{table}[!hbt]
   \centering
   \caption{Accelerator parameters of the SPS at flat-top.}
   \begin{tabular}{l c r}
       \toprule
       \textbf{Parameter}         &    \textbf{Unit}      & \textbf{Value} \\
       \hline
           Circumference, $2\pi R$      &       m         &  $6911.55 $    \\ %[3pt]
        Beam energy, $E$ & GeV &$200$\\
         Main harmonic number, $h$         &           & $4620$        \\ %[3pt]
         % Transition Lorentz factor & & $6.1$ \\
         Main rf frequency, $f_{\text{rf}}$   &  MHz   &   $200.39$    \\
         rf voltage at main harmonic, $V_{\text{rf}}$ & MV & $4.5$ \\
                 Frequency of the $4^{\text{th}}$ harmonic rf system  &  MHz   &   $800$ \\
        rf voltage of the $4^{\text{th}}$ harmonic rf system & kV & 450\\
       \toprule
   \end{tabular}
   \label{paramet}
\end{table}
% Similar to Sec.~\ref{sec:brk}
\begin{figure}[!tbh]
    \centering
    \includegraphics[width=0.89\linewidth]{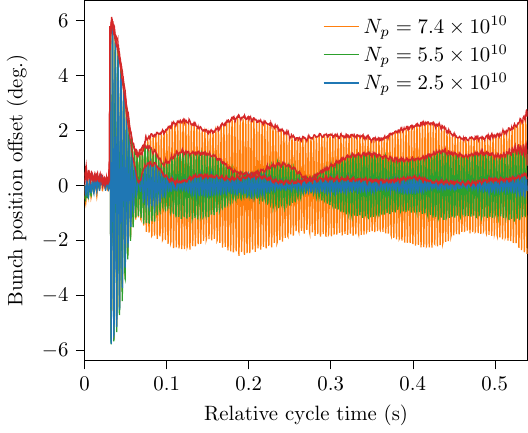}
    \caption{Bunch phase offset evolution at different intensities obtained by measuring the phase-loop error in single-harmonic rf ($200$~MHz). The red lines represent the envelopes of the traces computed using a Hilbert transform.}
    \label{fig:SRF_BE}
\end{figure}

\begin{figure}
\begin{subfigure}{.44\textwidth}
    \centering
    \includegraphics[width=\linewidth]{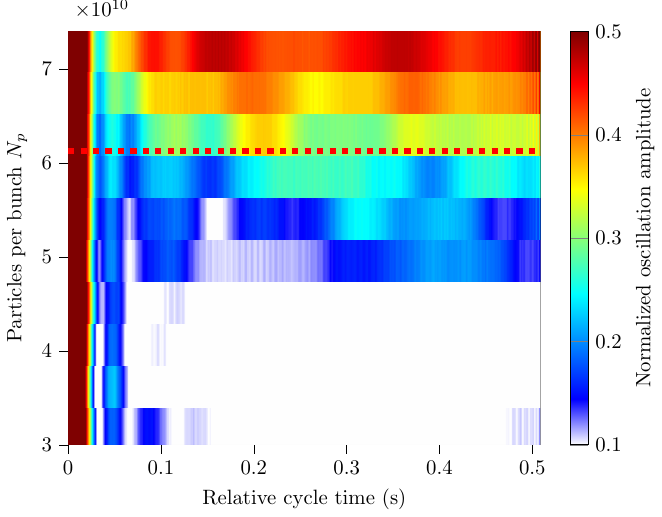}
    \caption{Measurement.}
    \label{fig:measurement}
\end{subfigure}
\begin{subfigure}{.44\textwidth}
    \centering
    \includegraphics[width=\linewidth]{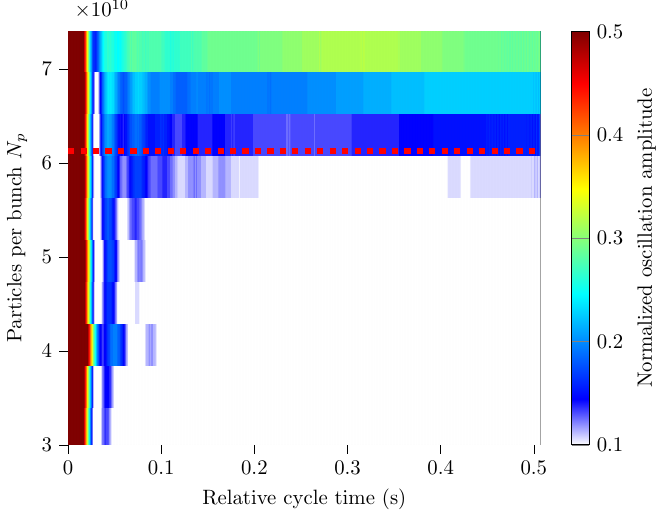}
    \caption{BLonD.}
    \label{fig:blond}
\end{subfigure}
\begin{subfigure}{.44\textwidth}
    \centering
    \includegraphics[width=\linewidth]{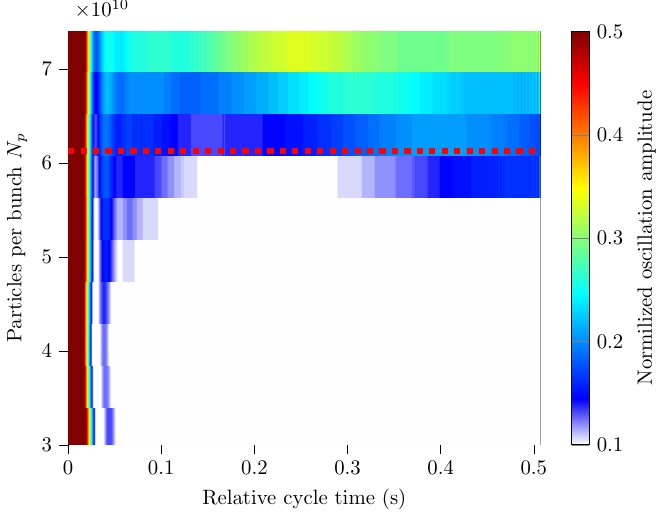}
    \caption{MELODY.}
    \label{fig:vkm} 
\end{subfigure}
\caption{Time evolution of the normalized bunch phase oscillation amplitude (color coding) in the SPS after a dipole excitation in the single-harmonic rf case. The measurement results (a) are compared with the outcome of BLonD~(b) and MELODY~(c) analysis for different intensities. The MELODY prediction of the LLD threshold at $N_{\text{th}}\approx 6.1 \times 10^{10}$ is shown in a red dashed line in all plots.}
\label{fig:SRF_color}
\end{figure}

\begin{figure}[!htbp]
% \begin{subfigure}{.24\textwidth}
    \centering
    \includegraphics[width=0.93\linewidth]{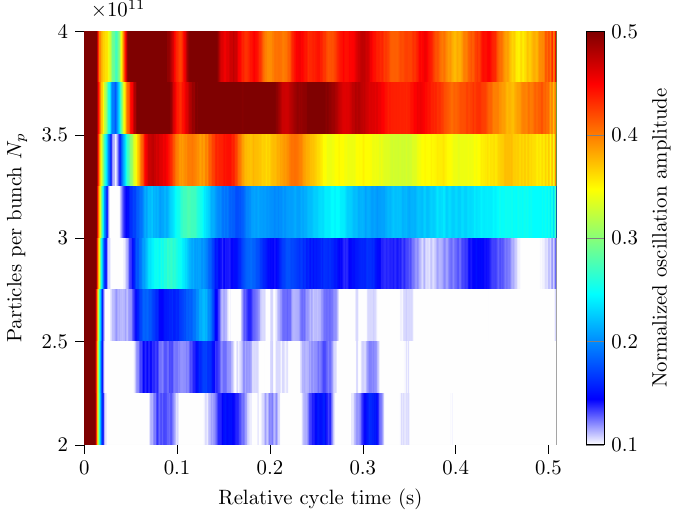}
    \caption{Measured time evolution of the bunch phase oscillation amplitude, normalized by the kick strength, in BSM for different intensities with voltage ratio $r_{V}=0.1$.}
    \label{fig:BSM}
\end{figure}
\begin{figure}[!htbp]
\begin{subfigure}{.42\textwidth}
    \centering
    \includegraphics[width=\linewidth]{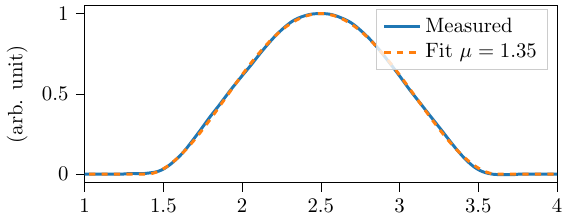}
    % \caption{SRF.}
    \label{fig:SRF_fit}
\end{subfigure}
\begin{subfigure}{.42\textwidth}
    \centering
    \includegraphics[width=\linewidth]{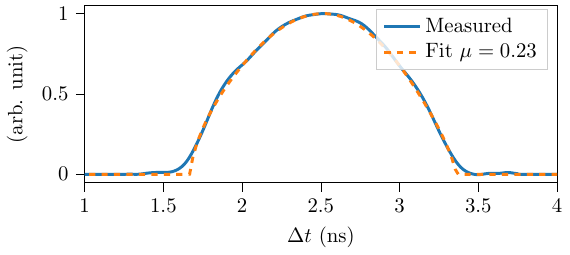}
    % \caption{BSM; $r=0.1$.}
    \label{fig:BSM_fit}
\end{subfigure}
\caption{Measured line density in single-harmonic rf~(top) and BSM (bottom) in the SPS at $200$~GeV. The binomial fit in Eq.~\eqref{eq:lambdabinomial} of the profile is depicted in orange.}
\label{fig:ld}
\end{figure}

To minimize the space charge contributions of the impedance and to enable a wider acquisition window due to the longer plateau, after injecting a single bunch from the PS at a total energy of $26$~GeV, the beam is accelerated to an energy of $200$~GeV. A few milliseconds after reaching the flat-top energy, the bunch is excited by a phase kick, inducing dipole oscillations within a rigid bucket. The beam phase loop~\cite{Baudrenghien:1415913} was disabled just before the phase excitation to prevent it from correcting the rf phase following the dipole excitation, which would rapidly damp the phase offset oscillation.
The evolution of the bunch offset is obtained by measuring the relative phase difference between the beam pickup signal and the sum of the cavity voltages~(i.e., phase-loop error~\cite{Spierer:2845762}), as shown in Fig.~\ref{fig:SRF_BE}.

Similar to Fig.~\ref{fig:kick}, following the phase excitation, beams with low intensity~(e.g., blue) rapidly lose coherence followed by subsequent slow decoherence. This indicates the typical behavior below the LLD threshold.
Above the threshold, phase oscillations persist (e.g., orange), with amplitudes dependent on the bunch intensity. The correlation of the residual oscillation with the intensity is attributed to crossing the LLD threshold. Below this intensity threshold, damping is dominant.

As a reference case, only the voltage contribution from the $200$~MHz system is considered for the moment. 
The measurements were performed with various bunch intensities ranging from~$3.0 \times 10^{10}$ to $7.0 \times 10^{10}$, in increments of~$\sim0.5\times 10^{10}$.

Figure~\ref{fig:SRF_color} shows the time evolution of the bunch phase oscillation amplitude (color coding), for different intensities. The oscillation amplitude from the turn-by-turn bunch offset was extracted using a Hilbert transform for each acqusition~\cite{hilbert}.
The measured profiles were fitted with Eq.~\eqref{eq:binomDistr} and subsequently used as input for simulations. 
In particular, Fig.~\ref{fig:blond} is derived with BLonD whereas, Fig.~\ref{fig:vkm} is obtained by means of MELODY expressing the rigid-dipole perturbation as a superposition of van Kampen modes. This allows us to analytically derive the evolution of the bunch oscillation as in~\cite{Karpov2021,alexahin1996landau}.
Both methods assume accelerator parameters according to Table~\ref{paramet}, including the updated SPS impedance model~\cite{SPSimpedance} after the impedance reduction campaign during the $2019-2021$ long shutdown. The two approaches exhibit similar behavior and agree well with the LLD threshold predicted by MELODY~(red dashed). However, the residual oscillation amplitudes depicted by measurement in Fig.\ref{fig:measurement} are larger than predicted by the simulations. The resulting LLD threshold predictions of $N_{\text{th}}^{\text{SRF}}=6.13\times 10^{10}$ for the single-harmonic rf case, slightly overestimate the measurements of about $25\%$. Overall, it agrees very well.

The SPS impedance model has been refined over many years through measurements and simulations of accelerator components~\cite{SPS1,SPS2}. However, accurately characterizing impedance behavior at high frequencies remains a challenge. As highlighted by Eq.~\eqref{eq:LLDth}, the LLD threshold is strongly influenced by the cutoff frequency of the impedance, $f_c$. Consequently, any inaccuracies in the impedance model or its cutoff frequency can significantly affect the threshold.

\begin{figure}[!tbh]
\begin{subfigure}{.483\textwidth}
    \centering
    \includegraphics[width=\linewidth]{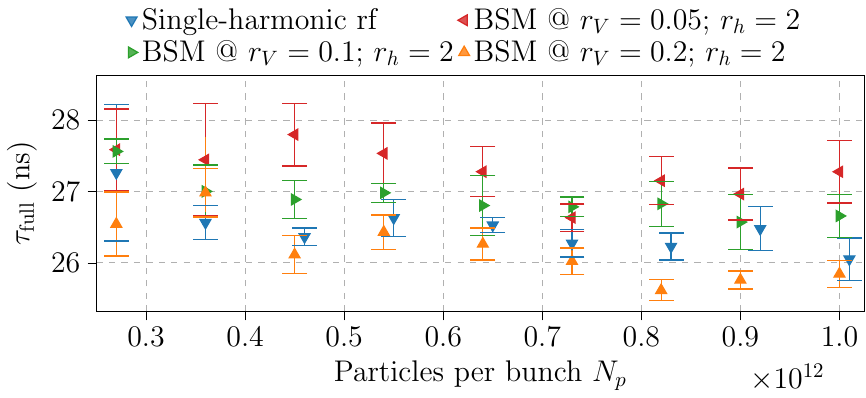}
    \end{subfigure}
\begin{subfigure}{.483\textwidth}
    \centering
    \includegraphics[width=\linewidth]{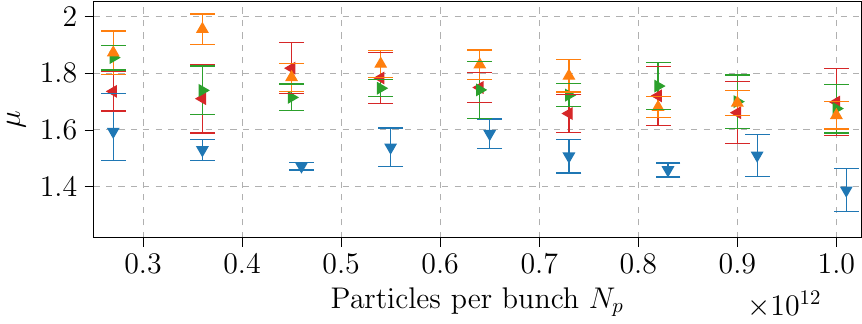}
    \end{subfigure}
 \caption{Mean values of the full bunch lengths and the tail form factor, $\mu$ in the PS and their $95\%$ confidence intervals in single-harmonic rf and BSM for different voltage ratios.}
\label{fig:bunchlength}
\end{figure}

\begin{figure*}
\begin{subfigure}{.495\textwidth}
  \centering
    \includegraphics[width=0.9\linewidth]{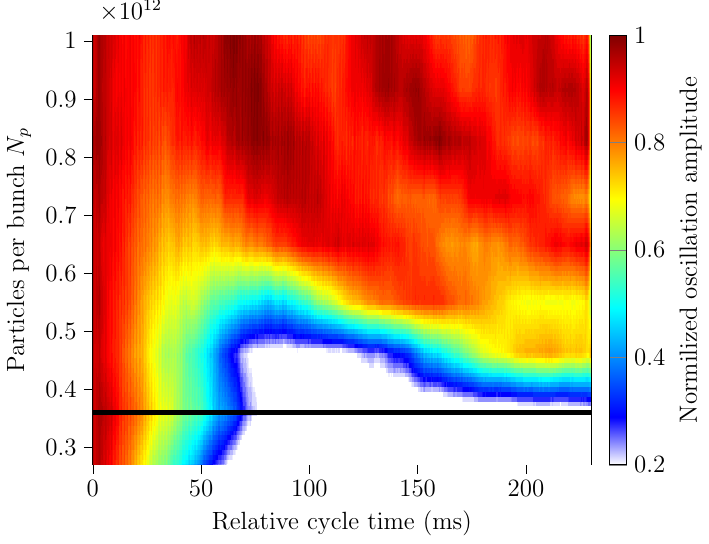}
  \caption{Single-harmonic rf.}
  \label{fig:sfig1}
\end{subfigure}%
\begin{subfigure}{.495\textwidth}
  \centering
    \includegraphics[width=0.9\linewidth]{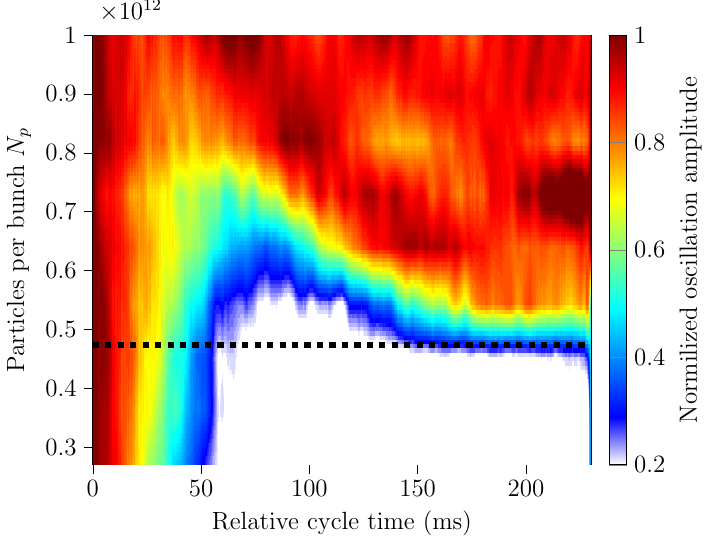}
  \caption{BSM with $r_{h}=2$; $r_{V}=0.05$.}
  \label{fig:sfig2}
\end{subfigure}

\begin{subfigure}{.495\textwidth}
  \centering
    \includegraphics[width=0.9\linewidth]{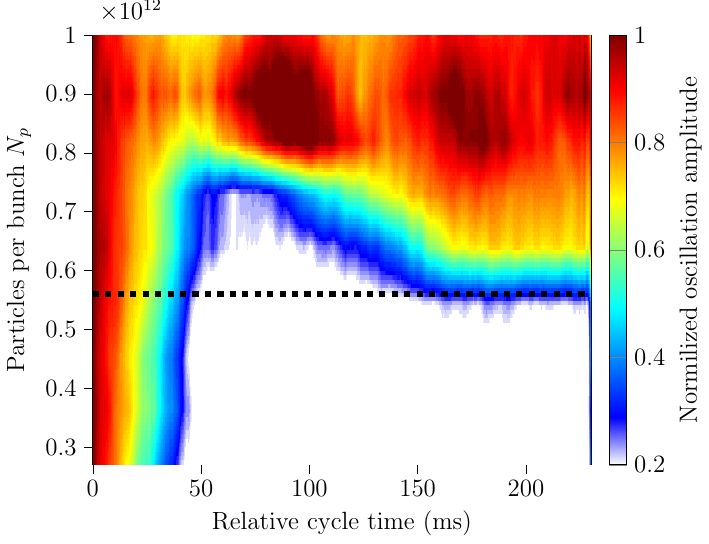}
  \caption{BSM with $r_{h}=2$; $r_{V}=0.1$.}
  \label{fig:sfig3}
\end{subfigure}
\begin{subfigure}{.495\textwidth}
  \centering
    \includegraphics[width=0.9\linewidth]{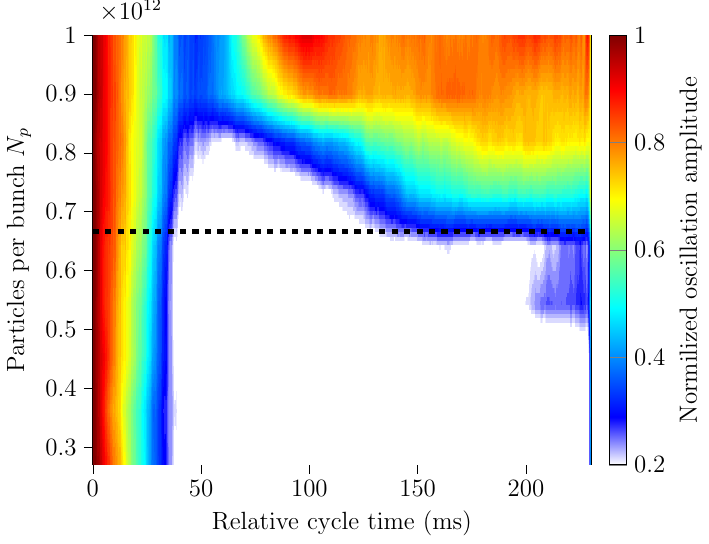}
  \caption{BSM with $r_{h}=2$; $r_{V}=0.2$.}
  \label{fig:sfig4}
\end{subfigure}
\caption{Time evolution of the normalized bunch phase oscillation amplitude (color coding) for different intensities after a dipole excitation in the PS.}
\label{fig:colorplot}
\end{figure*}

We now will include the voltage contribution from the $800$~MHz system in BSM ($200$~MHz and $800$~MHz rf voltage in phase at the bunch position in steady state).
Figure~\ref{fig:BSM} illustrates the measured time evolution of the normalized bunch phase oscillation amplitude, relative to the kick strength for a voltage ratio of $r_{V}=0.1$. A significant benefit in longitudinal stability can be observed when employing a double-harmonic rf system as compared to the single-harmonic rf case shown in Fig.~\ref{fig:measurement}. In particular, the BSM results in an increase of approximately $\sim6$ in terms of the LLD threshold. However, direct comparison is not straightforward due to the different distributions and longitudinal emittances. This is confirmed by the measured bunch profiles in Fig~\ref{fig:ld} with the corresponding binomial fit~(orange). Moving from the single-harmonic rf scenario (top) to the BSM case (bottom), a reduction in bunch length and a change in bunch shape (lower $\mu$) are visible. Furthermore, in the BSM scenario, a discrepancy is observed in the tail of the line density, indicating that the distribution is not entirely binomial.

\subsubsection{The LLD threshold in the PS}

To complement the studies we perform measurements in the PS, under very different beam conditions.
The large number of rf systems employed in the PS (i.e., $2.8-10$~MHz, $20$~MHz, $40$~MHz, $80$~MHz, and $200$~MHz) makes it an ideal accelerator for the LLD study. Furthermore, the low frequency of the rf systems features remarkable reproducibility and accurate control of both rf and beam parameters, as illustrated in Fig.~\ref{fig:bunchlength} for single-harmonic rf and BSM with different voltage ratios. It summarizes the mean values of the full bunch lengths and binomial tail form factor $\mu$, along with their $95\%$ confidence intervals, for each bunch intensity.

We will illustrate how the scaling factor in the LLD threshold Eq.~\eqref{eq:LLDth}, can be employed to evaluate the relative change of the LLD threshold at different rf settings.
The PS rf parameters are summarized in Table~\ref{parametPS}.

\begin{table}[!hbt]
   \centering
   \caption{Main PS parameters for LHC-type beams~\cite{LHCInjectors} at flat-top.}
   \begin{tabular}{l c r}
        \toprule 
       \textbf{Parameter}         &    \textbf{Unit}      & \textbf{Value} \\
       \hline
           Circumference, $2 \pi R$      &       m         &  $628.32 $    \\ %[3pt]
        Beam energy, $E_0$ & GeV &$26$\\
         Fundamental harmonic number, $h$         &           & $21$        \\ %[3pt]
         % Transition Lorentz factor & & $6.1$ \\
         Fundamental rf frequency, $f_{\text{rf}}$   &  MHz   &   $10$    \\
         rf voltage at fundamental harmonic, $V_0$ & kV & 200 \\
         % Second harmonic rf frequency   &  $\text{MHz}$   &   $20$    \\
         % Fourth harmonic rf frequency   &  $\text{MHz}$   &   $40$    \\
         rf voltage at $2^{\text{nd}}$ harmonic ($20$~MHz)  &  kV   &   (up to) $40$    \\
         % rf voltage at $4^{\text{th}}$ harmonic ($40$~MHz) &  kV   &   (up to) $200$    \\
         % Main rf voltage  & kV  &  $200$\\
        \toprule
   \end{tabular}
   \label{parametPS}
\end{table}

Following the same technique as performed in the SPS, we injected and accelerated a single bunch to the maximum energy of $26$~GeV. The flat-top was chosen to minimize any contribution from space charge and to ensure a dominant inductive beam-coupling impedance above transition energy. A few milliseconds after reaching the flat-top energy, the second rf system ($20$~MHz or $40$~MHz) was activated. The bunch was then excited by a phase kick with all beam control loops disabled.
The evolution of the bunch profile with respect to a beam synchronous trigger is obtained. A burst of trigger pulses is generated each of them separated by an integer number of turns, starting just before the excitation.
The bunch position evolution can be derived by computing the centroid of the acquired profiles with respect to the trigger.

For the second-harmonic rf system case in BSM ($r_{h}=2$), Fig.~\ref{fig:colorplot} shows the time evolution of the bunch phase oscillation amplitude (color coding) after a phase kick for different intensities.
Passing from the single-harmonic rf (a) to the BSM (b-d), following the initial rapid decoherence, it is observed that bunch oscillations become undamped at higher intensities for larger voltage ratios. 

Assuming the LLD threshold in single-harmonic rf is $N_{\text{th}}\approx0.36\times10^{12}$~(black solid line in Fig.~\ref{fig:sfig1}) 
we can employ Eq.~\eqref{eq:LLDth2} to predict the relative change in the threshold for other configurations. Taking into account the small variation in the bunch length and the tail form factor $\mu$ (Fig.~\ref{fig:bunchlength}), we obtain $N_{\text{th}, r=0.05}^{\text{BSM}} = (0.48 \pm 0.09) \times 10^{12}$ (Fig.\ref{fig:sfig2}); $N_{\text{th}, r=0.10}^{\text{BSM}} = (0.56 \pm 0.05) \times 10^{12}$~(Fig.\ref{fig:sfig3}); $N_{\text{th}, r=0.20}^{\text{BSM}}  =( 0.67 \pm 0.07) \times 10^{12}$ (Fig.\ref{fig:sfig4}). Consequently, the analytical predictions (black dashed) are consistent with the measurements as shown in Fig.~\ref{fig:colorplot}.

As far as the fourth harmonic rf system ($r_{h}=4$, higher harmonic rf frequency at $40$~MHz) is concerned, the measurements have been performed at the same bunch length as with $r_{h}=2$, again covering different voltage ratios. However, no LLD has been observed in the same range of intensities. This behavior is expected since already for a voltage ratio of $r_{V}=0.05$, using Eq.~\eqref{eq:LLDth2}, an LLD threshold increase by a factor of almost $4$ is predicted. It would therefore be well beyond the intensity reach for the given bunch length in the PS.
\begin{figure}[!tbh]
    \centering
    \includegraphics[width=0.85\linewidth]{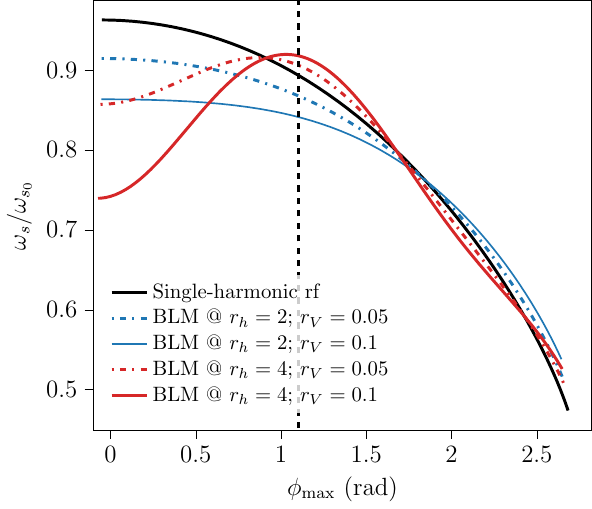}
    \caption{Synchrotron frequency distribution normalized to the small-amplitude synchrotron frequency, $\omega_{s0}$ as a function of the maximum phase deviation of the particle. The blue lines denote the BLM for $r_{h}=2$ and the red lines for $r_{h}=4$, when the intensity is $N_p=0.25\times 10^{12}$. The corresponding bunch length to $\tau_{\text{full}}=35$~ns, is highlighted with a black dashed line.}
    \label{fig:BLM_Synch}
\end{figure}

\begin{figure}[!tbh]
    \centering
    \includegraphics[width=0.9\linewidth]{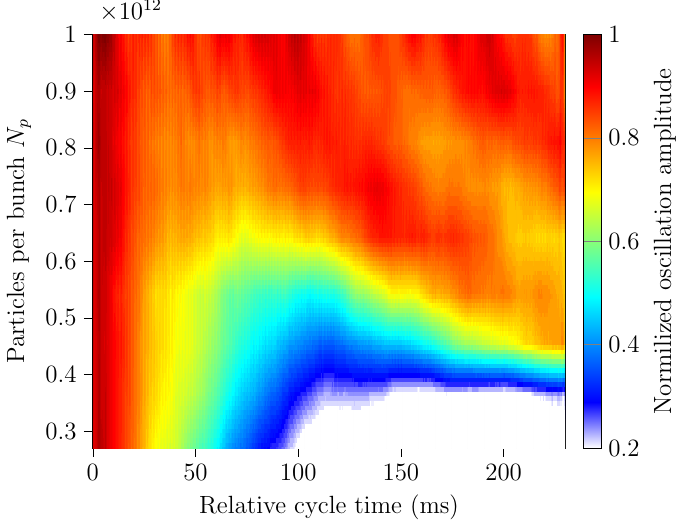}
    \caption{Time evolution of the normalized bunch offset amplitude (color coding) in BLM for $r_{h}=2$, $r_{V}=0.05$ and $\tau_{\text{full}}=35$~ns.}
    \label{fig:BLM}
\end{figure}

It has been shown that Landau damping can be significantly affected by the presence of zero derivatives of the synchrotron frequency distribution $d{\omega_s}/d{\phi}$. As illustrated in Fig.~\ref{fig:BLM_Synch} for $r_{h}=4$ (red), this inflection point ($d{\omega_s}/d{\phi}=0$) of the synchrotron frequency distribution is well below the current bunch length ($\tau_{\text{full}}=35$~ns), leading to the vanishing LLD threshold. This phenomenon was observed experimentally when the bunch position oscillation was persistently undamped also for the lowest possible bunch intensity~($N_{\text{th}}=0.27\times10^{12}$).
 With $r_h=2$ and $r_{V}=0.05$, the synchrotron frequency distribution, plotted in Fig.~\ref{fig:BLM_Synch} (blue, dash-dotted), is similar to the single-harmonic rf case without any significant enhancement in synchrotron frequency spread. This observation is consistent with the result presented in Fig.~\ref{fig:BLM} in which the addition of the higher-harmonic rf system does not bring benefits for the LLD threshold (with respect to~Fig.\ref{fig:sfig1}).
 
\section{CONCLUSION}
The loss of Landau damping (LLD) in synchrotrons is a critical condition that can lead to beam quality degradation and particle loss. It occurs when a coherent mode emerges from the incoherent band of particle oscillation frequencies. In the longitudinal plane, the LLD threshold can, in principle, be raised by reducing longitudinal impedance. However, increasing the synchrotron frequency spread through a multi-harmonic rf system is often much more effective. The present work expands the recent studies on longitudinal LLD threshold studies to the important case of double-harmonic rf systems with inductive impedance above transition energy (or capacitive impedance below). This scenario is relevant for the majority of high~(low) energy synchrotrons.

Refined analytical estimates for the synchrotron frequency distribution allowed the derivation of an analytical expression for the LLD threshold in the bunch shortening mode, where both rf systems are in phase at the bunch position in a non-accelerating bucket. The equation is obtained from the Lebedev equation, for binomial particle distributions, which covers most of the realistic longitudinal bunch distributions in proton accelerators.
Consistent with the single-harmonic rf case, the threshold is inversely proportional to the impedance cutoff frequency, $f_c$ when $f_c\gg 1/\tau_{\mathrm{full}}$, where $\tau_{\mathrm{full}}$ is the full bunch length. In line with past studies, the higher harmonic rf system significantly increases the LLD threshold. We have demonstrated that it scales with the factor $(1+r_{V} r_{h}^3)/(\mu^2+\mu)$, where $r_{V}$, $r_{h}$, and $\mu$ represent the voltage ratio, the harmonic number ratio, and tail form factor of the binomial distribution.
The analytical equation for the LLD threshold has been benchmarked with a semi-analytical approach using the code MELODY, showing excellent agreement, particularly for short bunches. Further validation was conducted by comparing MELODY with the macroparticle tracking simulation code BLonD. Both codes, based on different approaches to study longitudinal beam dynamics, return remarkably consistent results.
The non-monotonic dependence of the LLD threshold with respect to the bunch length, observed in past analyses, was clarified by the difference in the threshold definition, although it was based on the kick response. We showed that extremely slow decoherence after a phase kick can be observed in bunch shortening mode for intensities below the real LLD threshold, which remains a monotonic function of the bunch length.

The second relevant configuration examined is the bunch lengthening mode operating with both rf systems in counter-phase at the bunch position. While this mode initially appears to offer a higher synchrotron frequency spread, it presents significant limitations. Specifically, if the maximum of the synchrotron frequency distribution does not correspond to the maximum amplitude of the particle oscillation, the LLD occurs already at zero intensity.
Furthermore, given the high sensitivity of the bunch lengthening mode to the relative phase, $\Phi_2$, between the two rf systems, an extensive parameter scan was conducted to assess the LLD threshold in hybrid configurations. For instance, with a cutoff frequency of an order of magnitude above the rf frequency, $f_c=10 f_{\mathrm{rf}}$, a gain of factor three with respect to the bunch shortening mode was observed for $\Phi_2\approx 2 \pm \pi$. Nonetheless, the margin of error in the relative phase is very tight, presenting significant challenges for operating in this regime. This will be investigated in the future.

The study presents a straightforward beam-based measurement technique to study the LLD threshold experimentally. A single bunch is longitudinally excited by a phase kick to observe the decoherence and possibly undamped bunch offset oscillation in a rigid bucket due to the LLD. An extensive measurement campaign was performed in two different synchrotrons under different bucket filling conditions. At the SPS, we first considered the single-harmonic rf case, as a reference study and subsequently, we moved to the configuration of bunch shortening mode. The significant threshold increase for double-harmonic rf configurations has been demonstrated. The results were then compared with MELODY and BLonD, showing a remarkable agreement between the numerical predictions and the measurements, with only a $25\%$ discrepancy.
The residual difference might require further refinement in the broad-band part of the current SPS beam coupling impedance model and the assessment of potential noise excitation from the rf system.

To complement the study, we moreover performed measurements in the PS
employing double-harmonic rf systems with frequency ratios of two and four.
The measurements were mainly devoted to validate the scaling factor of the analytically derived equation for the LLD threshold. 
The findings showed a very good agreement with the predictions, and a longitudinal stability gain for different double-harmonic rf configurations was proved in bunch shortening mode for a harmonic number ratio of $r_{h}=2$. However, for a fourth harmonic rf system, the expected LLD threshold was so high that the intensity reach prevented the observation of LLD. As expected from theory, in bunch lengthening mode, instead, the bunches were long enough to remain undamped.

A similar analysis can also be devoted to the case of inductive impedance below transition energy (or capacitive impedance above), which will be the subject of further work.

\section{ACKNOWLEDGMENTS}
The authors would like to thank Simon Albright, Salvatore Energico, Alexandre Lasheen, and Danilo Quartullo for their valuable input and insightful discussions.

\nocite{*}

\bibliography{bib}

\end{document}